%% file: lw_qspec_v3_arxiv.tex
\documentclass[12pt]{article}
\usepackage{amsmath}
\usepackage{graphicx}
\usepackage{enumerate}
\usepackage{natbib}
\usepackage{amsmath}
\usepackage{url} 

\newcommand{\blind}{0}

\input{myfonts}

\usepackage{float}
\usepackage{caption}
\usepackage{subcaption}
\usepackage{booktabs}

\def\spacingset#1{\renewcommand{\baselinestretch}%
{#1}\small\normalsize} \spacingset{1}

 \DeclareMathSymbol{,}{\mathpunct}{operators}{"2C}

\begin{document}

\if0\blind
{
  \title{\bf  Quantile Fourier Transform, Quantile Series, and Nonparametric Estimation 
  of Quantile Spectra}
  \author{Ta-Hsin Li\footnote{Formerly affiliated with IBM T. J. Watson Research Center. 
  Email: {\sc thl024@outlook.com}} }
  \date{April 3, 2025}
  \maketitle
} \fi

\if1\blind
{
  \title{\bf  Quantile Fourier Transform, Quantile Series, and Nonparametric Estimation of Quantile Spectra} 
  \maketitle
} \fi

\begin{abstract}
Quantile spectra were  recently  introduced through trigonometric quantile regression as an alternative to the conventional power spectra for quantile-frequency analysis (QFA) of time series. As bivariate functions of frequecy and quantile level, quantile spectra have found useful application in a number of scientific and engineering disciplines including financial time series analysis and signal processing. In this paper, a nonparametric  method is proposed for estimating quantile spectra as bivariate functions of frequency and quantile level. This method is based on the quantile discrete Fourier transform (QDFT), derived from trigonometric quantile regression, and the resulting quantile series (QSER) as the inverse Fourier transform of the QDFT. The construction of the QSER enables an application of the lag-window (LW) approach to quantile spectrum estimation in a way similar to
conventional spectrum estimation. In addition, a post-processing step of smoothing across quantiles
is employed to further reduce the statistical variability of the LW estimator when the underlying spectrum 
varies smoothly with respect to the quantile level. This paper provides the results of a simulation study to evaluate the proposed method.

\vspace{1in}

\end{abstract}

\newpage
\spacingset{1.9} 

\section{Introduction}

The concept of quantile spectra and cross-spectra was introduced in Li (2008; 2012; 2014)  through
trigonometric quantile regression.  Let  
$\{ y_{j,t} \}$ $(j=1,\dots,m)$ be stationary processes with univariate  marginal distribution functions 
$F_{j}(y) := \Pr\{ y_{j,t} \le y\}$ and density functions $\dot{F}_j(y) > 0$ $(j=1,\dots,m)$. Let $F_{jj',\tau}(y,y') := \Pr\{ y_{j,t} \le y, \, y_{j',t-\tau} \le y'\}$  $(j,j'=1,\dots,m)$ be the lag-$\tau$ bivariate distribution functions and $\phi_{jj',\tau}(y,y') := \Pr\{ (y_{j,t} - y) \, (y_{j',t-\tau}-y' )<0 \}$ $(j,j'=1,\dots,m)$ 
be the  lag-$\tau$ bivariate level-crossing rates  $(\tau=0,\pm 1,\dots)$. The matrix of quantile 
spectra and cross-spectra, or simply the quantile spectrum,  of these processes at a quantile level $\al \in (0,1)$
 can be written as $\bS(\om,\al) : = [S_{jj'}(\om,\al)]_{j,j'=1}^m$, where
\eqn
S_{jj'}(\om,\al) := \eta_j(\al) \eta_{j'}(\al) \sum_{\tau = -\infty} ^\infty r_{jj'}(\tau,\al) \exp(-i\om \tau )
\quad (0 \le \om <2\pi).
\label{S}
\eqqn
In this expression, $\eta_j(\al) :=  \sqrt{\al(1-\al)} \, \kappa_j(\al)$, $\kappa_j(\al) := 1/\dot{F}_j(F_j^{-1}(\al))$, and
\eq
r_{jj'}(\tau,\al)  & := & 1- \frac{1}{2\al(1-\al)} \, \phi_{jj',\tau} \big(F_j^{-1}(\al) , F_{j'}^{-1}(\al) \big)  \\
&=& \frac{1}{\al(1-\al)} \, \Big\{ F_{jj',\tau}\big(F_j^{-1}(\al),F_{j'}^{-1}(\al)\big) - \al^2 \Big\}.
\eqq
The quantile spectrum  is  analogous to the conventional power spectrum in the sense that the latter is the asymptotic mean of the conventional periodogram and cross-periodogram metrix
whereas the former is the asymptotic mean of the quantile periodogram and cross-periodogram matrix constructed from trigonometric quantile regression. In this analogy, $\eta_j(\al)$ takes place of the standard deviation and $r_{jj'}(\tau,\al)$ takes place of the ordinary 
autocorrelation function.

Because  $\{ r_{jj'}(\tau,\al): j,j'=1,\dots,m\}$ coincides with the ordinary autocorrelation function 
of the indicator processes $\{ {\cal I}(y_{j,t} \le F_j^{-1}(\al) )\}$ $(j=1,\dots,m)$, the quantile spectrum 
in (\ref{S}) is a scaled version of the conventional spectrum of the indicator processes considered in  Davis and Mikosch (2009), Hagemann (2013), Dette et al.\ (2015), Barun\'{i}k and Kley (2019). While both spectra are representations of serial dependence in the frequency domain, the quantile spectrum in (\ref{S}) has the advantage 
of carrying the information about the univariate marginal distributions through 
the scaling function $\eta_j(\cdot)$. This information is discarded in the conventional spectrum of  indicator processes.

An advantage of the conventional spectrum of indicator processes is that it can be estimated 
straightforwardly by standard spectral estimation techniques. Estimating the quantile 
spectrum in (\ref{S}) is more challenging. Being the asymptotic mean of the quantile periodogram and cross-periodogram matrix,  the quantile spectrum can be simply estimated by periodogram smoothing.
However, other side-used spectral estimation techniques, such as  the lag-window (LW) and  
autoregression (AR) estimators, are not so easily applicable due to a lack of suitable time-domain 
data associated with the quantile periodogram. 
In this paper, we introduce the quantile series (QSER) as such needed time-domain
surrogate and thereby enable the application of these alternative spectral estimation techniques for quantile spectrum estimation. In this paper, we focus exclusively on the LW technique which results in  a nonparametric estimator derived from  the QSER. The AR approach will be discussed elsewhere.

The conventional spectrum of indicator processes is typically treated as a univariate  function 
of frequency $\om$ for fixed  $\al$ like a conventional spectrum (e.g., Dette et al.\ 2015).
We regard the quantile spectrum  in (\ref{S}) as a bivariate function of $\om$ and $\al$ on 
$[0,2\pi) \times (0,1)$. Exploration of this bivariate function constitutes what we call quantile-frequency analysis or QFA, which has proven fruitful in a number of applications including financial and engineering time series  (Chen et al.\ 2019; Jim\'{e}nez-Var\'{o}n et al.\ 2024; Li 2020; 2021; 2023). In this paper, we do not intend 
to address further applications
 
The gist of the proposed method is as follows: First, we use the solutions of trigonometric quantile regression to construct what we call the quantile discrete Fourier transform or QDFT for each series $\{ y_{j,t}: t=1,\dots,n\}$ $(j=1,\dots,m)$ on a finite grid of quantile levels; then, for each quantile level, we compute the inverse Fourier transform of the QDFT to produce $m$ sequences in the time domain, which we call the quantile series (QSER);  and finally, we use the sample autocovariance function of the QSER, which we call the quantile autocovariance function (QACF), to construct the LW estimator of the quantile spectrum in (\ref{S}) by following the lag-window approach which is traditionally used to estimate the conventional  spectrum from the ACF. We further employ a smoothing 
procedure to smooth the LW estimates across quantiles with the aim of reducing statistical variability 
when the underlying spectrum is suitably smooth with respect to the quantile level. We call the resulting estimator 
the LW estimator with quantile smoothing or LWQS. The quantile smoothing step is proven effective in our simulation study.

In this paper, we focus on the description of the proposed method and 
use a simulation study to evaluate its performance.
We leave the theoretical exploration of sampling properties to future research. 
The remainder of this paper is organized as follows. 
In Section 2, we introduce the QDFT, QSER, and QACF. 
In Section 3, we describe the LWQS estimator.
In Section 4, we present the results of the simulation study. Concluding remarks are given in Section 5.
Additional results of the simulation study in  Appendix I. A summary of R functions 
for the proposed method is provided in Appendix II.

\section{Quantile Fourier Transform and Quantile Series}

Given a data record $\{ y_{j,t}: t=1,\dots,n\}$  of length $n$, 
let $\om_v := 2\pi v/n$ $(v=0,1,\dots,n-1)$ be the $n$ Fourier frequencies. 
For each $\om_v \notin \{0,\pi\}$, consider the following 
trigonometric quantile regression solution at quantile level $\al \in (0,1)$:
\eqn
\lefteqn{
\{ \hat{\beta}_{1,j}(\om_v,\al),
\hat{\beta}_{2,j}(\om_v,\al), \hat{\beta}_{3,j}(\om_v,\al)\} } \notag \\
& := &  \operatorname*{argmin}_{\beta_1,\beta_2,\beta_3 \in \bbR} \ \sum_{t=1}^n
\rho_\al \big(y_{j,t} - \beta_1 - \beta_2 \cos(\om_v t) - \beta_3 \sin(\om_v t) \big),
\label{qr1}
\eqqn
where $\rho_\al(y) :=  y \, (\al - {\cal I}(y \le 0))$ is the objective function
of quantile regression (Koenker 2005, p.\ 5). In addition, for $\om_v = \pi$ (i.e., $v=n/2$ when $n$ is even), let
\eqn
\{ \hat{\beta}_{1,j}(\pi,\al),
\hat{\beta}_{2,j}(\pi,\al)\}
& := & \operatorname*{argmin}_{\beta_1,\beta_2 \in \bbR} \ \sum_{t=1}^n
\rho_\al \big(y_{j,t} - \beta_1 - \beta_2 \cos(\pi t) \big), \label{qr2} \\
 \hat{\beta}_{3,j}(\pi,\al) & := & 0, \notag 
\eqqn
and for $\om_v = 0$ (i.e., $v=0$), let
\eqn
\hat{\beta}_{1,j}(0,\al)
& := & \arg\min_{\beta_1 \in \bbR}\sum_{t=1}^n \rho_\al \big(y_{j,t} - \beta_1 \big),  \label{qr3} \\
\hat{\beta}_{2,j}(0,\al) & := & \hat{\beta}_{3,j}(0,\al) \ := \ 0.
\notag
\eqqn
Based on these trigonometric quantile regression solutions, we define the quantile discrete Fourier transform (QDFT) 
of $\{ y_{j,t}: t=1,\dots,n\}$ at quantile level $\al$ as
\eqn
Z_j(\om_v,\al)  := 
\left\{ 
\begin{array}{ll}
n \, \hat{\beta}_{1,j}(0,\al) & v=0, \\
n \, \hat{\beta}_{2,j}(\pi,\al) & v = n/2 \mbox{ (if $n$ is even)}, \\
(n/2) \{ \hat{\beta}_{2,j}(\om_v,\al) - i \, \hat{\beta}_{3,j}(\om_v,\al) \} & \text{otherwise}.
\end{array} \right.
\label{qdft}
\eqqn
 This definition of QDFT is motivated by the fact that the ordinary DFT  can be constructed in the same way by replacing $\rho_\al(y)$ with the objective function $y^2$ of least-squares regression.

It is easy to see that the sequence $\{ Z_j(\om_v,\al) :v=0,1,\dots,n-1\}$  is conjugate symmetric:
\eqn
 Z_j(\om_v,\al) = Z_j^*(\om_{n-v},\al) \quad (v=1,\dots,\lfloor (n-1)/2 \rfloor).
\label{sym}
\eqqn
Therefore, in order to compute the QDFT, one only need to solve 
the quantile regression problems (\ref{qr1})--(\ref{qr2})  for $\om_v \in (0,\pi]$, i.e., for
$v=1,\dots,(n-1)/2$ when $n$ is odd and  $v=1,\dots,n/2$ when $n$ is even; the conjugate symmetry property
provides the values of QDFT for the remaining frequencies. Linear programming algorithms
such as those implemented by the function {\tt rq} in the R package `quantreg' (Koenker 2005) can be employed
to compute the quantile regression solutions  efficiently. 

Based on the QDFT in (\ref{qdft}), the quantile periodograms  and 
cross-periodograms of the  $m$ series $\{ y_{j,t}:t=1,\dots,n\}$ $(j=1,\dots,m)$ 
at quantile level $\al$ can be written as
\eqn
Q_{jj'}(\om_v,\al) := n^{-1} Z_j(\om_v,\al)  \, Z_{j'}^*(\om_v,\al) \quad (v=0,1,\dots,n-1).
\label{qcper}
\eqqn
This expression of quantile periodograms and cross-periodograms in terms of the QDFT is analogous to the definition of conventional periodograms and cross-periodograms 
in terms of the ordinary DFT (Brockwell and Davis 1992, p.\ 443).
Under suitable conditions Li (2012; 2014, p.\ 557), it can be shown
that $Q_{jj'}(\om,\al) \dlim \zeta_j \zeta_{j'}^*$ for fixed $\om \in (0,\pi)$ and $\al \in (0,1)$, 
where $\bmzeta := [\zeta_1,\dots,\zeta_m]^T$ is complex Gaussian with mean  vector $\0$ and covariance matrix 
$\bS(\om,\al)$ defined by (\ref{S}). This  is similar to the result for the conventional periodogram where
$\bS(\om,\al)$ is replaced by the ordinary spectrum (Brockwell and Davis 1992, p.\ 446).

For each $j=1,\dots,m$, we can also  compute the inverse Fourier transform of  the QDFT: 
\eqn
x_{j,t}(\al) :=\frac{1}{n} \sum_{j=0}^{n-1}  Z_j(\om_v,\al) \exp(i t\om_v) \quad (t=1,\dots,n).
\label{qser}
\eqqn
We call this sequence the quantile series (QSER) of $\{ y_{j,t}: t=1,\dots,n \}$ at quantile level $\al$. Note that the QSER  is a real-valued time series due to  (\ref{sym}). Also note that the sample mean of the QSER, $\bar{x}_j(\al) := n^{-1} \sum_{t=1}^n x_{j,t}(\al)$, coincides with $\hat{\beta}_{1,j}(0,\al)$, which is nothing but the $\al$-quantile of $\{ y_{j,t} : t=1,\dots,n \}$, because $Z_j(0,\al) = n \, \hat{\beta}_{1,j}(0,\al)$ by definition. 
The QSER have the desired property that their conventional periodograms and cross-periodograms coincide
with the quantile periodograms and cross-periodograms $\{ Q_{jj'}(\om_v,\al): v=0.1,\dots,n-1\}$. 
Therefore, they can be used as a time-domain surrogate for quantile spectral estimation using conventional techniques.

\section{Lag-Window Spectral Estimator}

In matrix notation, let $\bx_t(\al) := [x_{1,t}(\al),\dots,x_{m,t}(\al)]^T$. 
Then, the sample autocovariance function (ACF) 
of the QSER in (\ref{qser}) is given by
\eqn
\hat{\bGam}(\tau,\al) := \frac{1}{n} \sum_{t=\tau+1}^{n} ( \bx_t(\al) - \bar{\bx}(\al)) \, 
(\bx_{t-\tau}(\al) - \bar{\bx}(\al))^T \quad  (\tau =0,1,\dots,n-1),
\label{qacf}
\eqqn
where $\bar{\bx}(\al) := n^{-1} \sum_{t=1}^n \bx_t(\al) = [\bar{x}_{1}(\al),\dots,\bar{x}_m(\al)]^T$.
We call $\hat{\bGam}(\tau,\al)$ in (\ref{qacf}) the sample quantile autocovariance  function (QACF)  at quantile level $\al$.
It is easy to show that the usual relationship between the ordinary ACF and the ordinary periodogram (Brockwell and Davis 1992, p.\ 443) holds true for the QACF and the QPER, i.e., 
\eqn
\bQ(\om_v,\al) = \sum_{|\tau|< n} \hat{\bGam}(\tau,\al) \exp(-i \om_v \tau) \quad (v = 1,\dots,n-1),
\label{qacf2qcper}
\eqqn
where $\bQ(\om_v,\al)  := [Q_{jj'}(\om_v,\al)]_{j,j'=1}^m$ and $\hat{\bGam}(-\tau,\al) := \hat{\bGam}(\tau,\al)^T$
$(\tau=1,\dots,n-1)$. In the remainder of this paper,  we redefine $\bQ(0,\al)$ as $\0$ instead of 
$n \, \bar{\bx}(\al) \bar{\bx}^T(\al)$ so that the expression in
(\ref{qacf2qcper}) remains valid for $v=0$. We also extend the quantile periodogram periodoically 
with period $2\pi$ so that the expression in
(\ref{qacf2qcper}) holds for all $v=0,\pm 1,\dots$.

In light of this relationship, we take the conventional lag-window (LW) approach  (Priestley 1981, p.\ 434)
and propose the following estimator for the quantile spectrum $\bS(\om,\al)$:
\eqn
\hat{\bS}_{\rm LW}(\om,\al)
:= \sum_{|\tau| <n} h(\tau/M) \, \hat{\bGam}(\tau,\al) \exp(-i \om \tau),
\label{LW}
\eqqn
where $\hat{\bGam}(\tau,\al)$ is the QACF given by (\ref{qacf}) and 
$h(\cdot)$ is an even and piecewise continuous function, satisfying 
$h(0)=1$, $|h(x)| \le 1$ for all $x$, and $h(x)=0$ for $|x| > 1$. An example of the lag window
is the Tukey-Hanning window (Priestley 1981, p.\ 443) 
\eqn
h(x) := \frac{1}{2} ( 1 + \cos(\pi x))  {\cal I}(|x| \le 1).
\label{Hanning}
\eqqn 
In the spectral case where $h(x) \equiv 1$ and $M=n-1$, the LW estimator $\hat{\bS}_{\rm LW}(\om_v,\al)$
becomes the quantile periodogram $\bQ(\om_v,\al)$ $(v=0,1,\dots,n-1)$ according to (\ref{qacf2qcper}).

To justify $\hat{\bS}_{\rm LW}(\om,\al)$ in (\ref{LW}) as an estimator of $\bS(\om,\al)$ in (\ref{S}), consider 
\eq
y_{j,t}(\al) :=  F_j^{-1}(\al) + \kappa_j(\al) \, u_{j,t}(\al),
\eqq
where
\eqn
u_{j,t}(\al) := \al - {\cal I}( y_{j,t} \le F_j^{-1}(\al)).
\label{qc}
\eqqn
 Let $\bGam(\tau,\al) := [\gam_{jj'}(\tau,\al)]_{j,j'=1}^m$ denote the ACF 
of $\by_t(\al) := [y_{1,t}(\al),\dots,y_{m,t}(\al)]^T$. 
Because the processes $\{ u_{j,t}(\al) \}$ $(j=1,\dots,m)$ are stationary 
with mean 0, variance $\al(1-\al)$, and autocorrelation function $\{r_{jj'}(\tau,\al): \tau=0, \pm 1, \dots\}$, it
follows that
\eq
\gam_{jj'}(\tau,\al) = \eta_j(\al) \, \eta_{j'}(\al) \, r_{jj'}(\tau,\al).
\eqq 
When all entries of the ACF are absolutely summable over $\tau$, we have 
\eqn
\bS(\om,\al) = \sum_{\tau=\infty}^\infty \bGam(\om,\al) \exp(-i \om \tau) \quad \om \in [0,2\pi).
\label{S2}
\eqqn
In other words, the quantile spectrum $\bS(\om,\al)$ is nothing but the ordinary spectrum 
of the stationary process $\{\by_t(\al)\}$.

On the other hand, under suitable conditions (Wu 2007; Li 2012), the  quantile regression coefficients in (\ref{qr1})--(\ref{qr3}) have the Bahadur-type representations
\eq
 \hat{\beta}_{1,j}(0,\al) & = & F_j^{-1}(\al) + \kappa_j(\al)  \, n^{-1}  \sum_{t=1}^n  u_{j,t}(\al) + o_P(n^{-1/2}), \\
  \hat{\beta}_{2,j}(\pi,\al) & = & \kappa_j(\al)  \, n^{-1}  \sum_{t=1}^n  u_{j,t}(\al) \cos(\pi t) + o_P(n^{-1/2}), \\
 \hat{\beta}_{2,j}(\om_{v},\al) & = & 2 \kappa_j(\al)  \, n^{-1} \sum_{t=1}^n u_{j,t}(\al) \cos(\om_{v} t) + 
o_P(n^{-1/2}) \quad 
 \om_{v} \notin \{ 0,\pi\}, \\
 \hat{\beta}_{3,j}(\om_{v},\al) & = & 2 \kappa_j(\al) \, n^{-1} \sum_{t=1}^n u_{j,t}(\al) \sin(\om_{v} t) 
+ o_P(n^{-1/2}) \quad \om_{v} \notin \{0,\pi\}.
\eqq
This implies that
\eq
x_{j,t}(\al) = y_{j,t}(\al) + e_{j,t}(\al)
\eqq
and $n^{-1} \sum_{t=1}^n \{e_{j,t}(\al) \}^2= o_P(1)$. Therefore, the QACF  $\hat{\bGam}(\tau,\al)$ in (\ref{qacf}) can be regarded as an estimate of $\bGam(\tau,\al)$ from certain noisy observations of $\{\by_t(\al)\}$.  Substituting this estimate in (\ref{S2}) together with
a suitable window $h(\cdot)$ leads to the LW estimator in (\ref{LW}).

By following the lines of conventional spectral estimation (Brockwell and Davis 1991, pp.\ 358--359),  
consider the spectral window
\eq
H(\om)  :=(2\pi)^{-1} \sum_{|\tau| \le M} h(\tau/M) \exp(-i \om \tau)
\eqq
and the continuous extention of the quantile periodogram
\eq
\bQ(\om,\al) := \sum_{|\tau| < n} \hat{\bGam}(\tau,\al) \exp(-i \om \tau)
\eqq
which concides with (\ref{qacf2qcper}) at all Fourier frequencies.
It can be shown that
\eq
\hat{\bS}_{\rm LW}(\om,\al)
& = & \int_{-\pi}^\pi H(\lam) \, \bQ(\om+\lam,\al) \, d\lam \\
& \approx & 
\sum_{|u| \le [n/2]} (2\pi/n) H(\om_u) \, \bQ(\om_{v+u},\al),
\eqq
where $\om_v$ is the nearest Fourier frequency to $\om$. In other words, 
the LW estimator can be viewed approximately as a type of periodogram smoother. 
See Priestley (1981, pp.\ 435--449) or Percival and Walden (1993, pp.\ 271--277) 
for discussions on different choice of lag windows.

For conventional spectral estimation, the LW technique yields a consistent estimator 
as $n \linfty$ if $M \linfty$ and $M/n \lzero$ (Brockwell and Davis 1991, p.\ 359; Priestley 1981, p.\ 457). This assertion is established by a calculation of the mean and variance of the LW estimator as an approximate 
periodogram smoother.  Due to a lack of closed-form solution, a direct calculation like this is not possible
for the quantile periodogram. However, we conjecture that a similar
consistency result should hold true for the LW estimator of the quantile spectrum. A definitive proof of this assertion remains an open problem for future research.

While the LW estimator in (\ref{LW}) is expected to work well for fixed $\al$, 
there are situations in which $\bS(\om,\al)$ changes smoothly with $\al$ and this
smoothness should be taken into account.
For example, $\bS(\om,\al)$ is a continuous function of $\al$ when (a) $\dot{F}_j(F_j^{-1}(\al))$ 
is continuous in $\al$ for all $j$, (b) $r_{jj'}(\tau,\al)$ is continuous in $\al$ for all $j$, $j'$, and $\tau$, and (c) 
$r_{jj'}(\tau,\al)$ is uniformly summable over $\tau$ for all $j$ and $j'$.
In such cases,  further improvement in estimation accuracy is expected if the smoothness 
is properly leveraged. This can be done, for example, by first evaluating the LW estimator 
on a finite grid of quantile levels $\{ \al_\ell: \ell=1,\dots,L) \subset [\ep, 1-\ep]$ for some small $\ep > 0$ 
and then applying a smoothing procedure to the resulting sequence $\{ \hat{\bS}_{\rm LW}(\om,\al_\ell): \ell=1,\dots,L\}$ for fixed $\om$. This estimator will be referred to as the  LW estimator 
with quantile smoothing, or  LWQS. It is a nonparametric alternative to the semiparametric estimators
considered in Chen et al.\ (2019) and Jim\'{e}nez-Var\'{o}n et al.\ (2024).

The proposed method assumes that the grid of quantile levels is given {\it a priori}\/ based on 
the need of application and other considerations. Ideally, the  design of the quantile grid should reflect the 
understanding or expectation of the general characteristics of the underlying spectrum as a function of $\al$: a fine grid is needed if the spectrum changes rapidly in $\al$, whereas a coarse grid is sufficient if the spectrum changes slowly in $\al$. It should also depend on the capability of the quantile smoothing procedure
to interplate between the quantiles. Like sample quantiles given by the order statistics, quantile regression in general has a 
finite number of distinct solutions which correspond to a finite number of quantile levels  (Portnoy 1991). 
These quantile levels can be produced numerically by the R function {\tt rq} in the `quantreg' package  (Koenker 2005, p.\ 303); they constitute the finest grid necessary for QFA. As a common practice, extreme quantiles (relative to the sample size) should be left out for special handling as they have different sampling
properties (Koenker 2005, p.\ 130; Davis and Mikosch 2009).

\section{ Simulation Study}

To demonstrate and justify the proposed estimation method, 
we employ two sets of simulated data with $m=2$ and $n=512$. We measure the accuracy of estimation by 
the Kullback-Leibler divergence
\eq
{\rm KLD} := 
\frac{1}{L \lfloor (n-1)/2 \rfloor} \sum_{\ell=1}^L
\sum_{v=1}^{\lfloor (n-1)/2 \rfloor} \bigg\{ \tr \big( \hat{\bS}(\om_v,\al_\ell) \bS^{-1}(\om_v,\al_\ell) \big)
- \log \frac{| \hat{\bS}(\om_v,\al_\ell)| }{ |\bS(\om_v,\al_\ell)| } - m \bigg\}.
\eqq
It is a nonnegative quantity which equals zero if and only if $\hat{\bS}(\om_v,\al_\ell) = \bS(\om_v,\al_\ell)$
for all $v$ and $\ell$. Being  closely related to Whittle's likelihood  (Whittle 1953), this spectral meaure has been 
used to quantify similarities for time series clustering and classification (Kakizawa,  Shumway, and Tanaguch  1998).

In our first set of simulated data, the first series, $\{ y_{1,t} \}$, is a nonlinear mixture of these components 
$\{ \xi_{1,t} \}$, $\{ \xi_{2,t} \}$,  and $\{ \xi_{3,t} \}$:
\eqn
\left\{
\begin{array}{lll}
z_t  & := & \psi_1(\xi_{1,t}) \times \xi_{1,t} + (1-\psi_1(\xi_{1,t})) \times \xi_{2,t}, \\
y_{1,t}  & := & \psi_2(z_t) \times z_t + (1-\psi_2(z_t)) \times \xi_{3,t},
\end{array}
\right.
\label{y1}
\eqqn
where $\psi_1(u) := 0.9 {\cal I}(u < -0.8) 
+ 0.2  {\cal I}(u > 0.8) + \{ 0.9 - (7/16) (u + 0.8)\} {\cal I}(|u| \le 0.8)$ and
$\psi_2(u) := 0.5 {\cal I}(u < -0.4) +  {\cal I}(u > 0.4) 
+ \{ 0.5 + (5/8) (u + 0.4)\} {\cal I}(|u| \le 0.4)$. 
The second series, $\{y_{2,t}\}$, is a delayed copy of $\{ \xi_{3,t} \}$:
\eqn
y_{2,t} := \xi_{3,t-10}.
\label{y2}
\eqqn
The three components  are 
zero-mean unit-variance  autoregressive (AR) processes, satisfying
\eq
\xi_{1,t} & = & a_{11} \, \xi_{1,t-1} + \ep_{1,t}, \\
\xi_{2,t} & = & a_{21} \, \xi_{2,t-1} + \ep_{2,t}, \\
\xi_{3,t} &= & a_{31} \, \xi_{3,t-1} + a_{32} \, \xi_{3,t-2} + \ep_{3,t},
\eqq
where $a_{11}  := 0.8$, $a_{21} := -0.7$, 
$a_{31} := 2r\cos(2\pi f_0)$ and $a_{32} := -r^2$ with $r=0.9$, $f_0 =0.2$, 
and where $\{\ep_{1,t} \}$, $\{\ep_{2,t} \}$, and  $\{\ep_{3,t} \}$ 
are mutually independent Gaussian white noise. In other words,
$\{ \xi_{1,t} \}$ is a low-pass series with spectral peak at frequency $0$, 
$\{ \xi_{2,t} \}$ is a high-pass series with spectral peak at frequency $\pi$, 
and $\{ \xi_{3,t} \}$ is a band-pass series with spectral peak at frequency $0.2 \times 2 \pi$. 
The mixing function $\psi_1(u)$ and $\psi_2(u)$ are designed to promote or reduce these spectral patterns 
at different quantile regions.

\begin{figure}[t]
\centering
\includegraphics[height=4.5in,angle=-90]{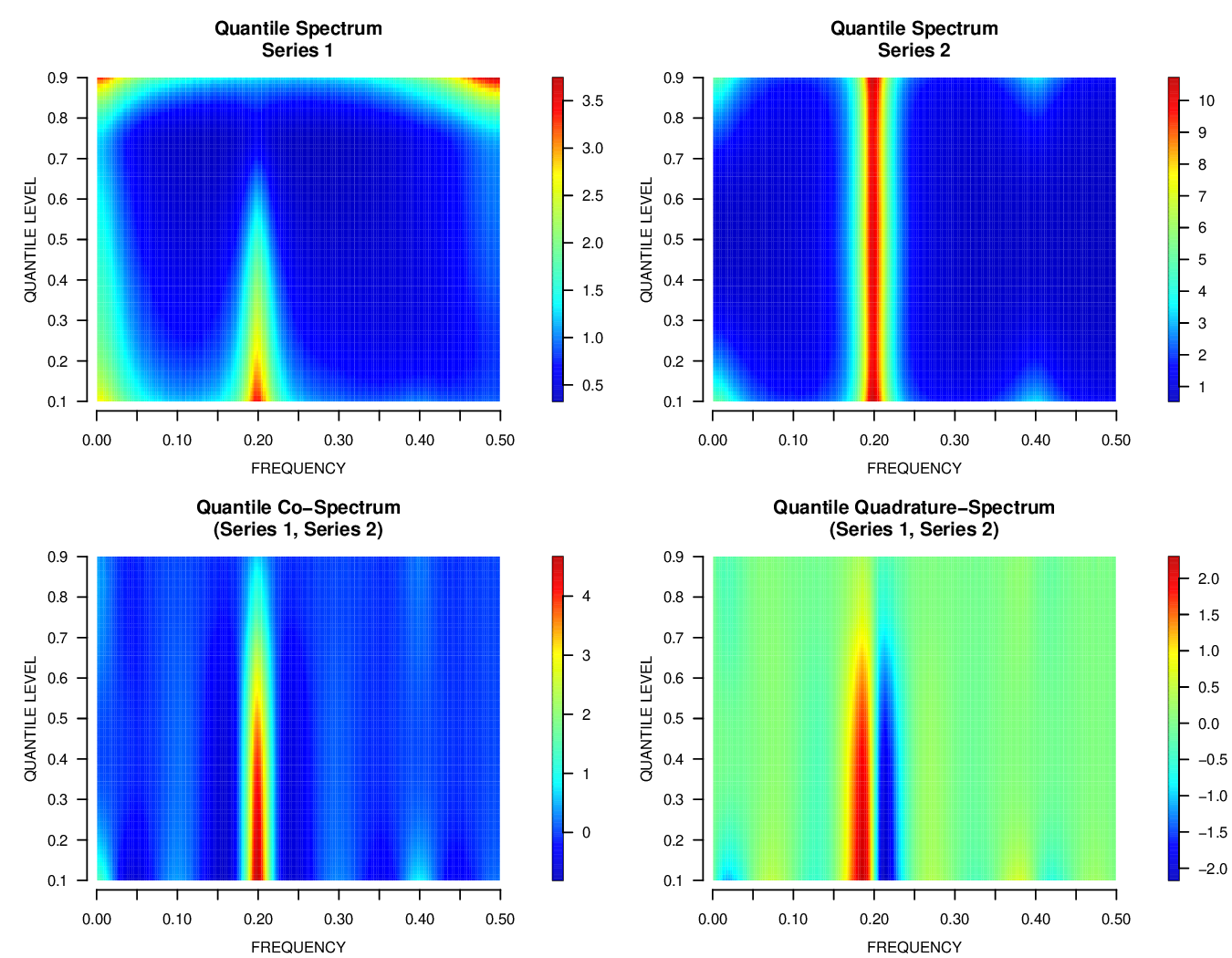}  
\caption{Quantile spectrum and cross-spectrum for the  mixture process  (\ref{y1})-(\ref{y2}). 
All spectra are shown as functions of  linear frequency
$f := \om/(2\pi) \in (0,0.5)$. } \label{fig:qspec}
\end{figure}

Figure~\ref{fig:qspec} shows the quantile spectrum and cross-spectrum of the mixture process  (\ref{y1})-(\ref{y2})
evaluated at $\om_v = 2\pi  v/512$ $(v=1,\dots,255)$ and $\al_\ell = 0.1 + 0.01 (\ell-1)$ $(\ell=1,\dots,81)$
($\{ \al_\ell \} = \{ 0.10,0.11,\dots,0.90\}$).  These spectra are computed as the ensemble mean of quantile periodograms and cross-periodograms from 5000  Monte Carlo runs. The cross-spectrum $S_{1,2}(\om,\al)$ is shown  by its  real and complex parts in the second row of Figure~\ref{fig:qspec}, which are 
known as co-spectrum and quadrature-spectrum, respectively,

\begin{figure}[p]
\centering
\includegraphics[height=5in,angle=-90]{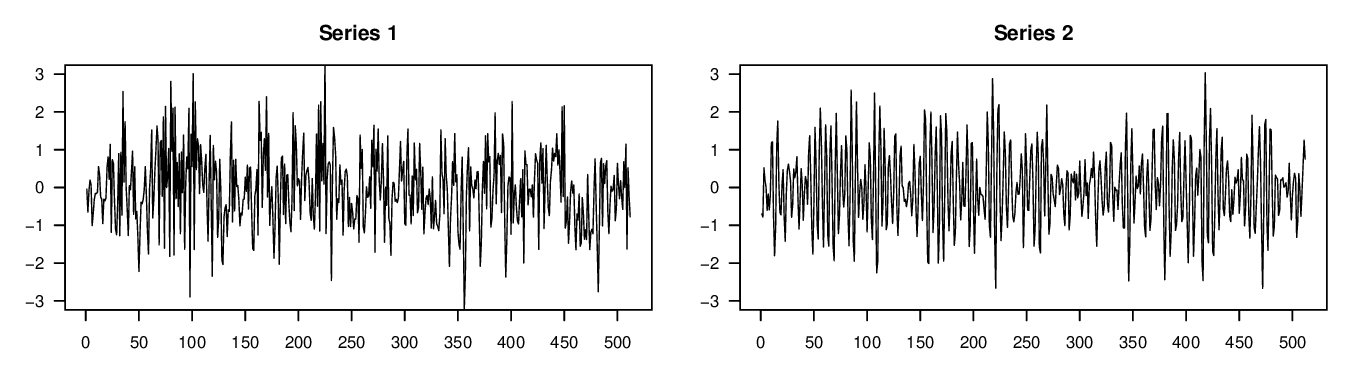} 
\caption{An example of simulated time series $(n=512)$ according to (\ref{y1}) and (\ref{y2}). }
\label{fig:ts}
\vspace{0.4in}
\includegraphics[height=4.5in,angle=-90]{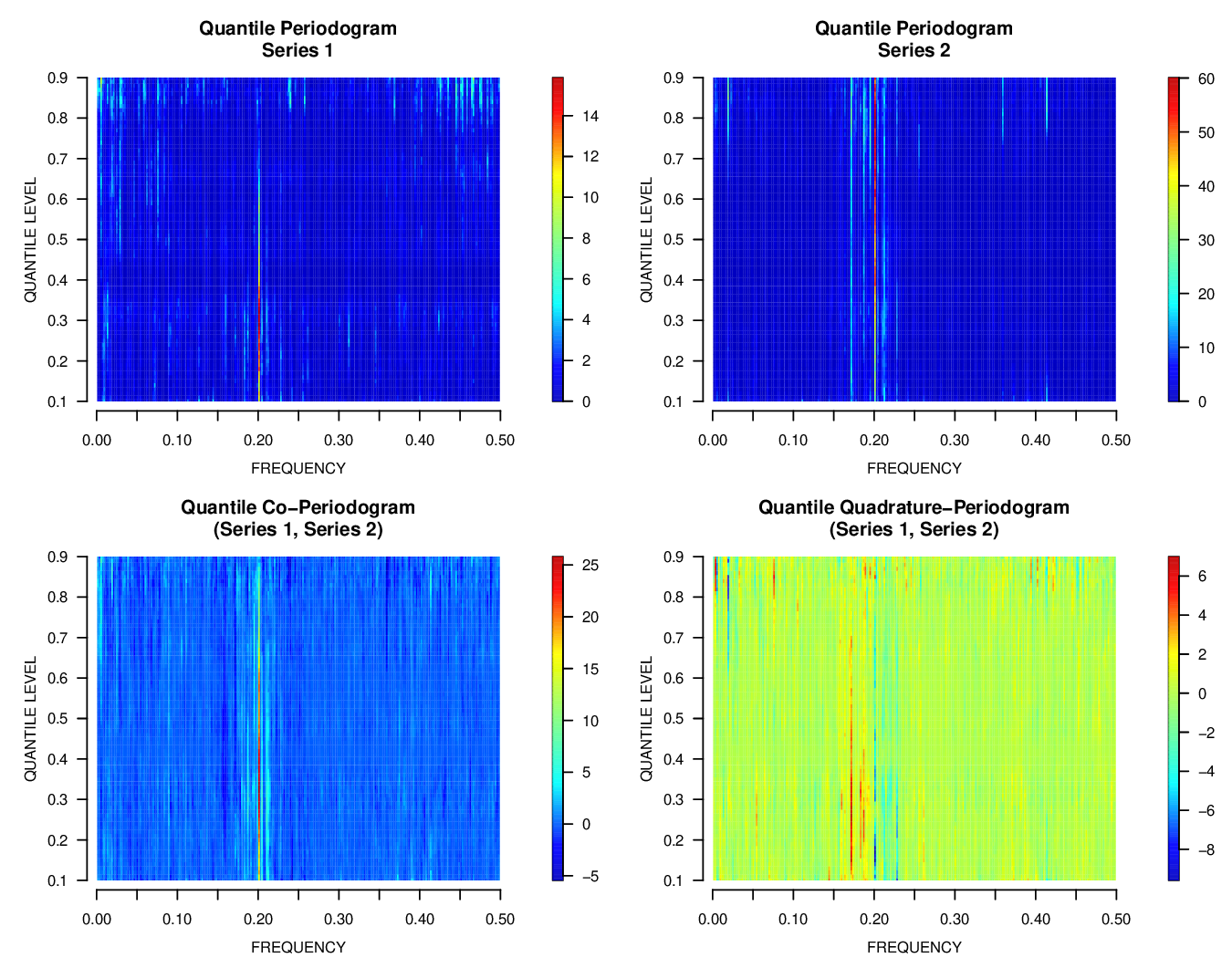}  
\caption{Quantile periodogram and cross-periodogram for the series shown in Figure~\ref{fig:ts}. }
 \label{fig:qcper}
\end{figure}

\begin{figure}[p]
\centering
\includegraphics[height=5in,angle=-90]{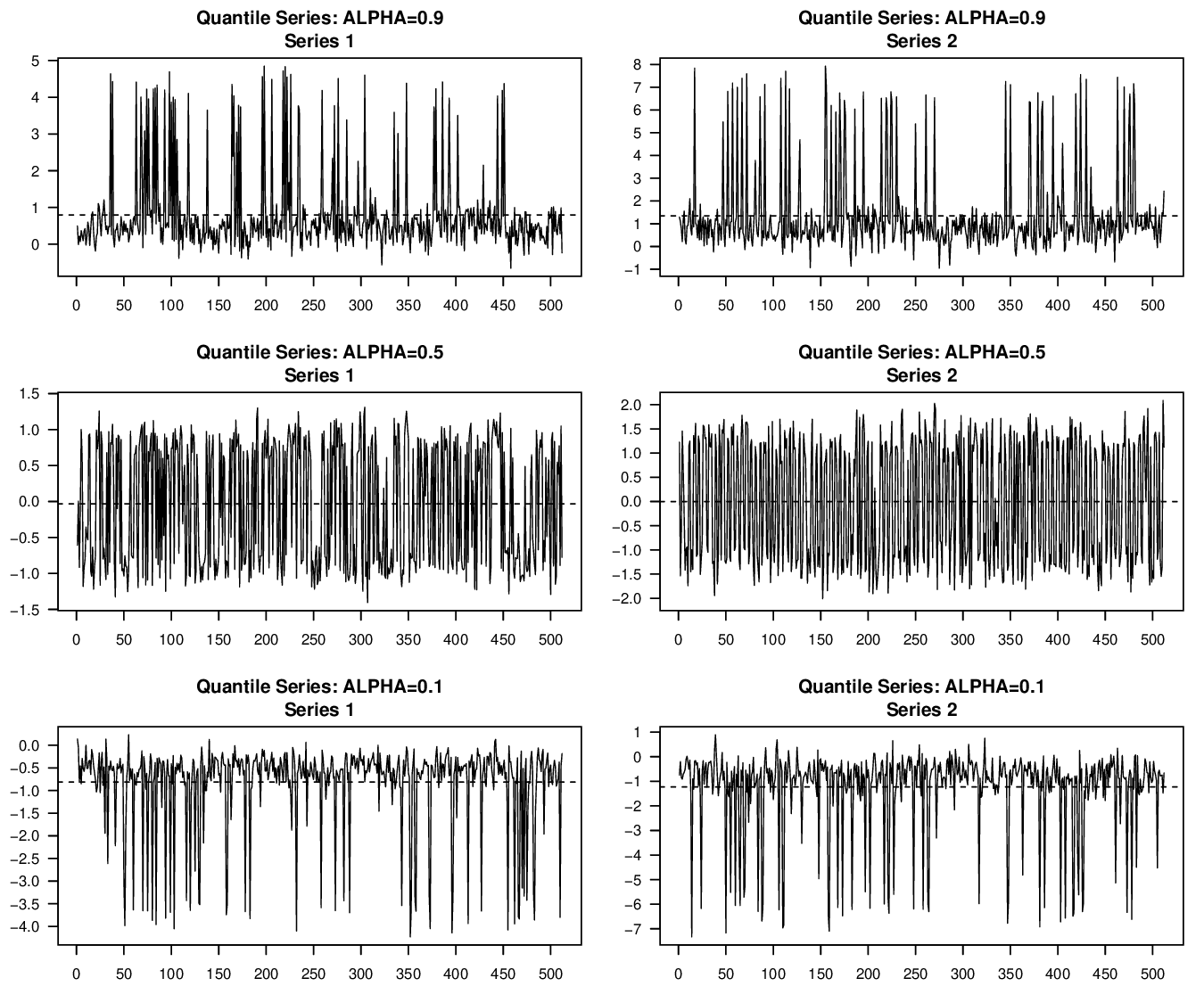} 
\caption{Time-series plot of quantile series for the series shown in Figure~\ref{fig:ts} at $\al =0.9$ (first row), $\al =0.5$ (second row), and $\al=0.1$ (third row).  
Dashed horizontal line depicts the sample mean of the quantile series. }
\label{fig:qser}
\end{figure}

\begin{figure}[p]
\centering
\vspace{0.5in}
\includegraphics[height=5in,angle=-90]{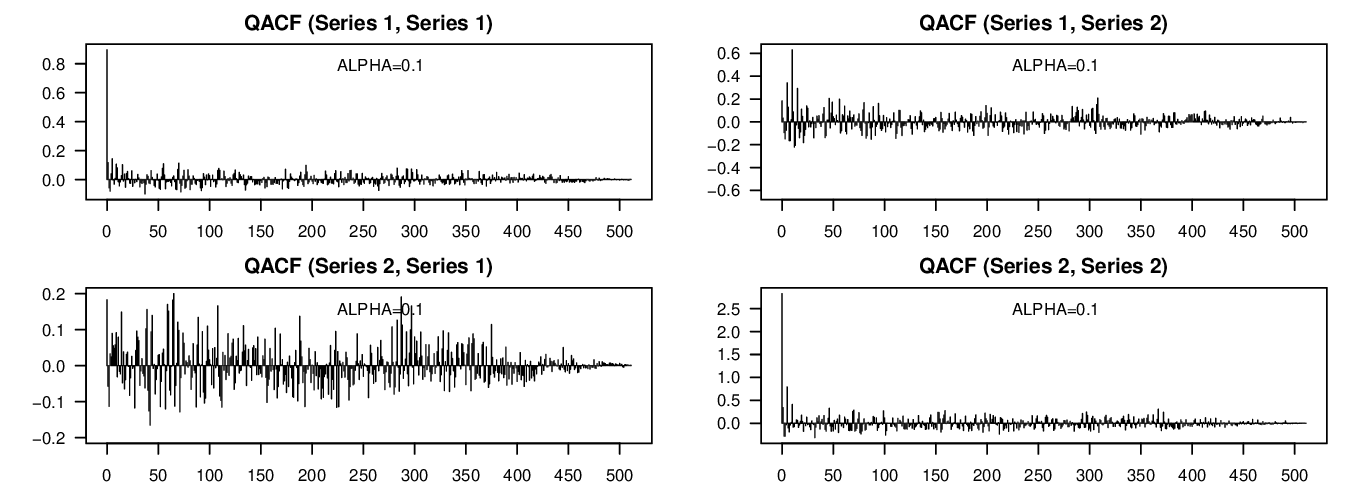}  \\[0.15in]
\includegraphics[height=5in,angle=-90]{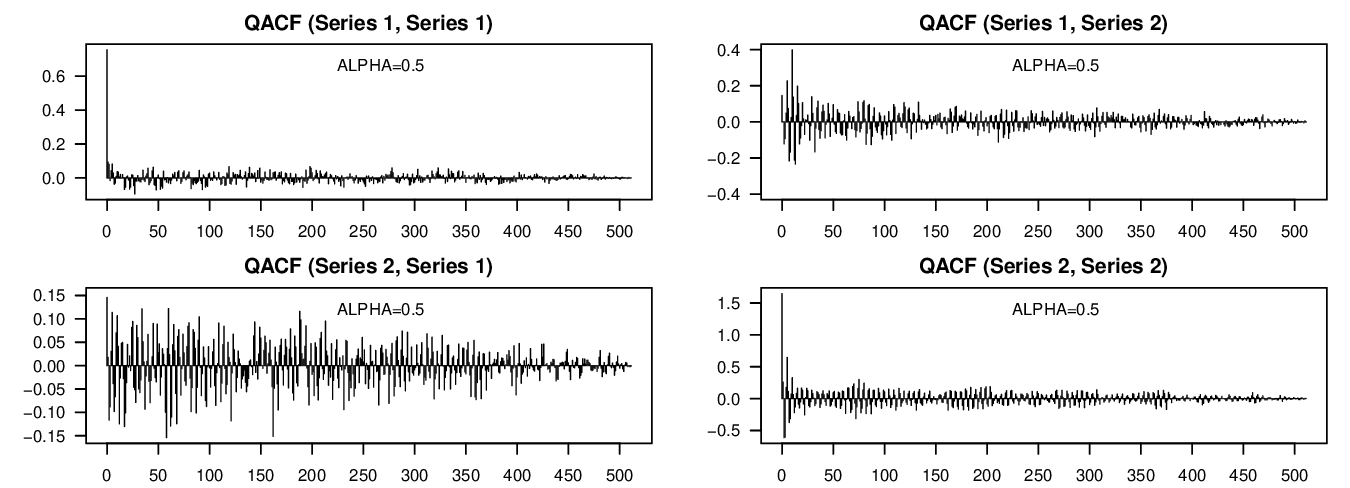}  \\[0.15in]
\includegraphics[height=5in,angle=-90]{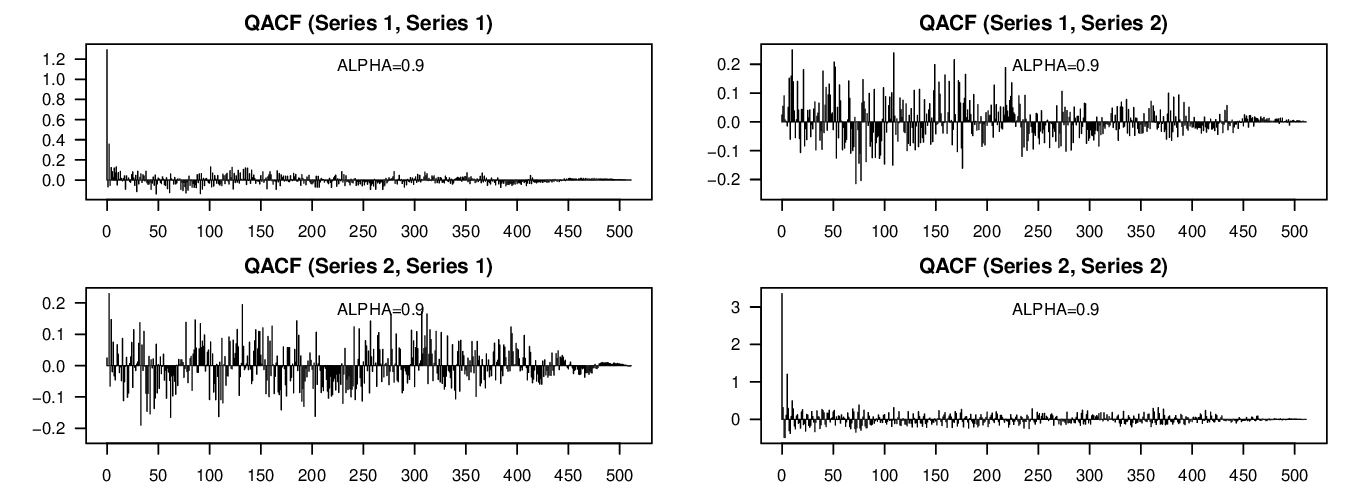}  \\[0.15in]
\caption{QACF of the series shown in Figure~\ref{fig:ts} at $\al =0.1$, $0.5$, and $0.9$. }
 \label{fig:qacf}
\end{figure}

\begin{figure}[p]
\centering
\includegraphics[height=4.5in,angle=-90]{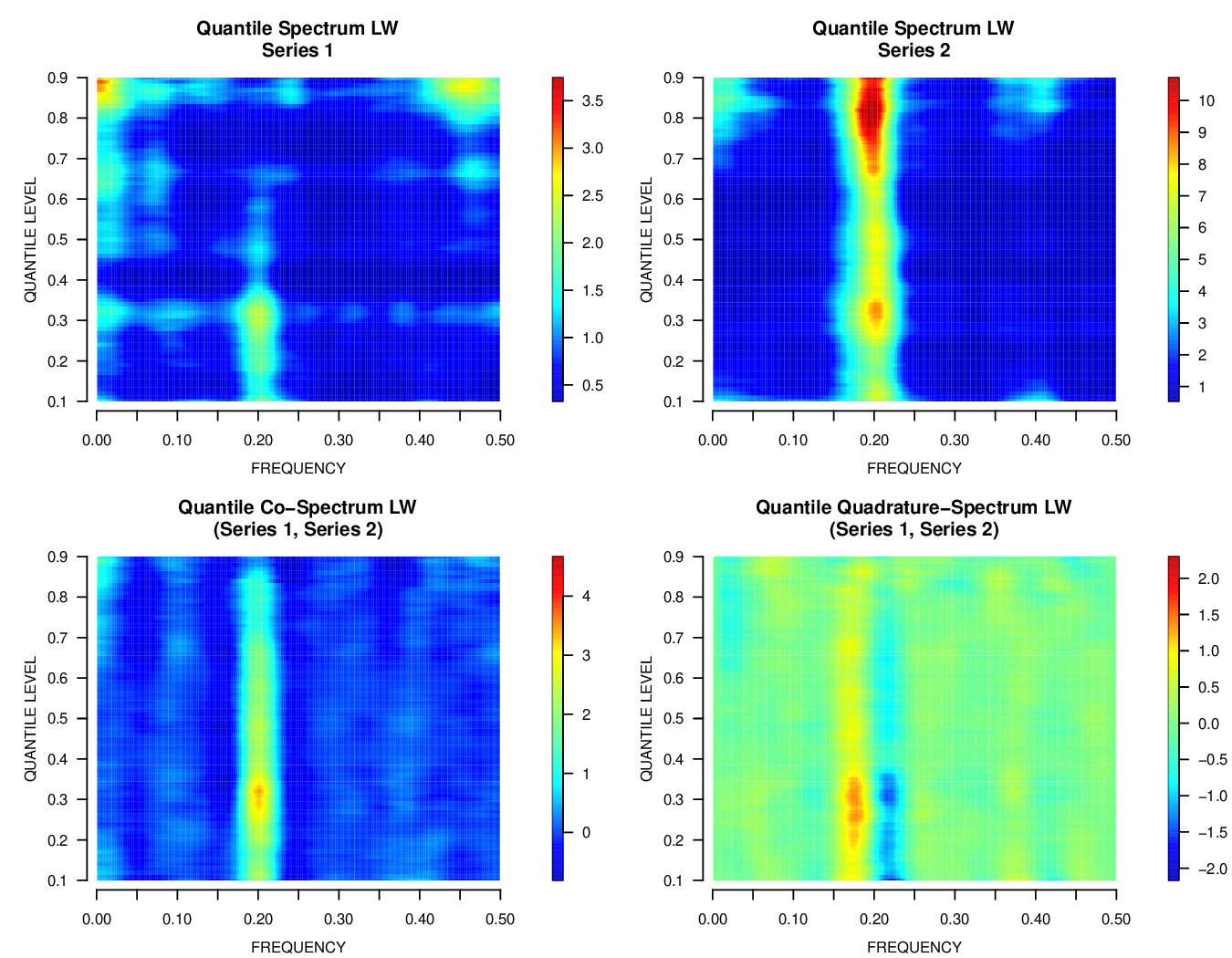}  
\caption{LW estimates of the quantile spectrum and cross-spectrum shown in Figure~\ref{fig:qspec} 
from the series shown in Figure~\ref{fig:ts}.
 KLD = 0.198.}
\label{fig:lw}
\centering
\includegraphics[height=4.5in,angle=-90]{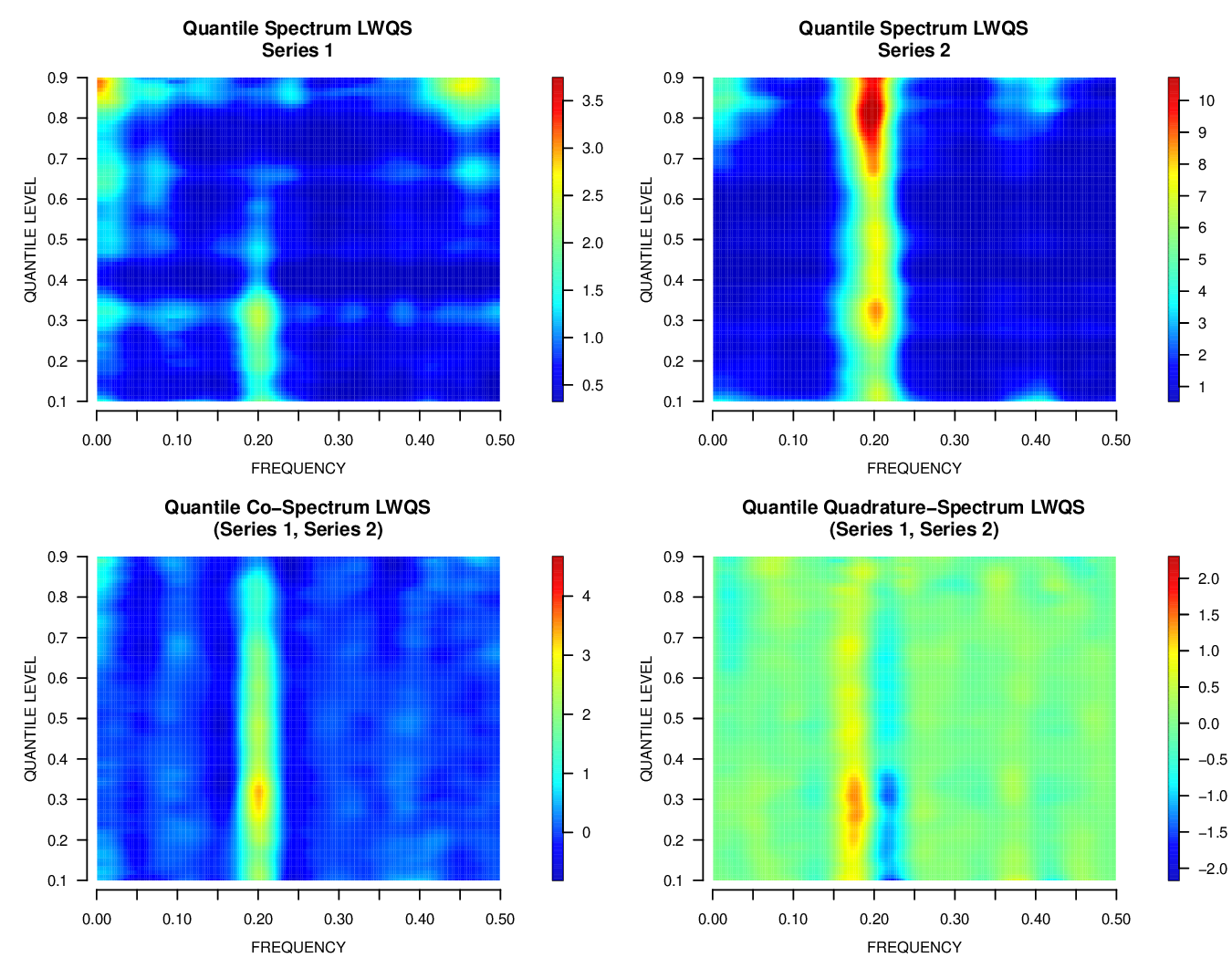}  
\caption{LWQS estimates of the quantile spectrum and cross-spectrum shown in Figure~\ref{fig:qspec}  from the series shown in Figure~\ref{fig:ts}
using {\tt smooth.spline} with GCV. KLD = 0.194. }
 \label{fig:lwqs}
\end{figure}

\begin{figure}[p]
\centering
\includegraphics[height=4.5in,angle=-90]{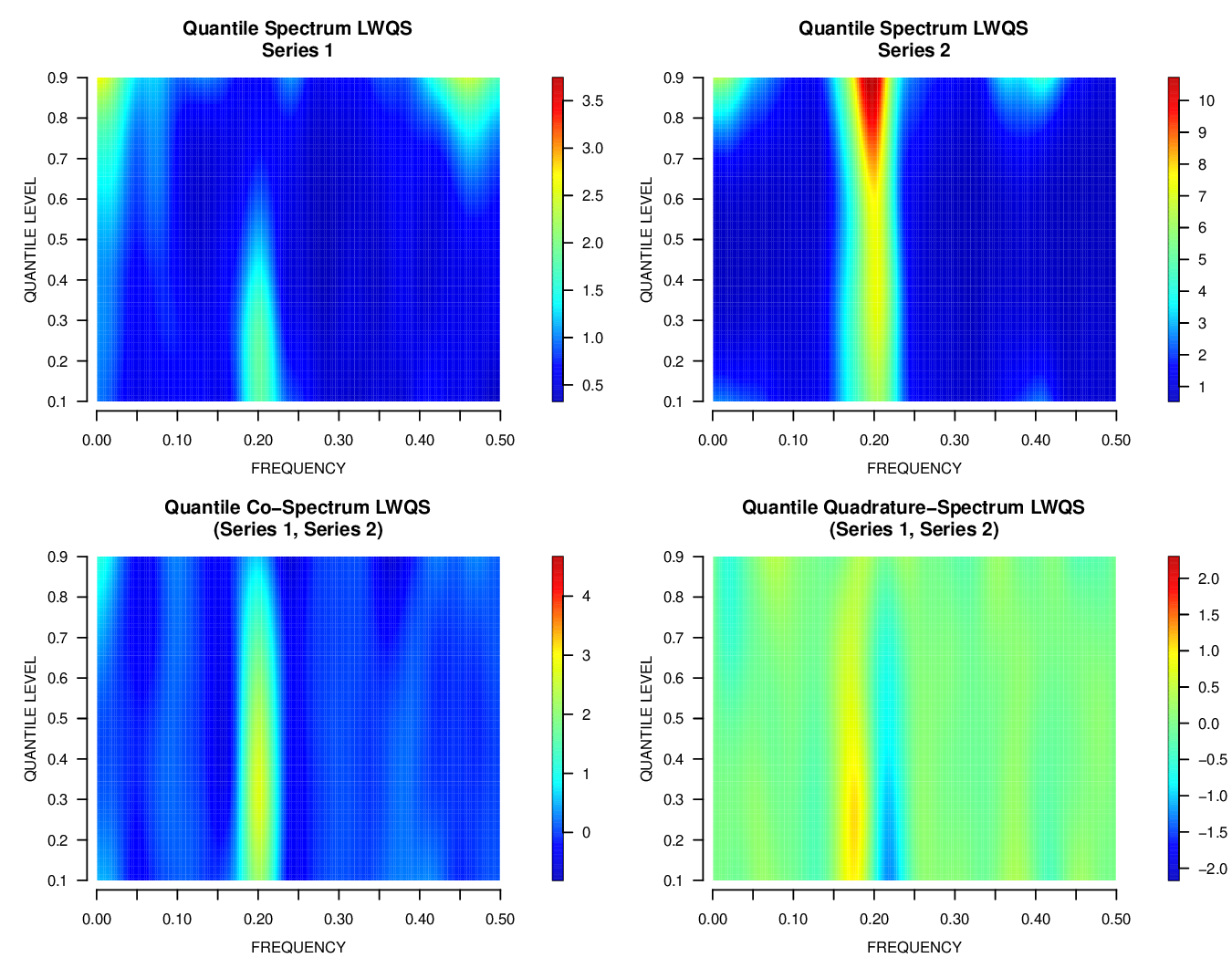}  
\caption{LWQS estimates of the quantile spectrum and cross-spectrum shown in Figure~\ref{fig:qspec}  for the series shown in Figure~\ref{fig:ts}  using {\tt smooth.spline} with {\tt spar} = 0.9.  KLD = 0.109.}
 \label{fig:lwqs2}
\centering
\includegraphics[height=4.5in,angle=-90]{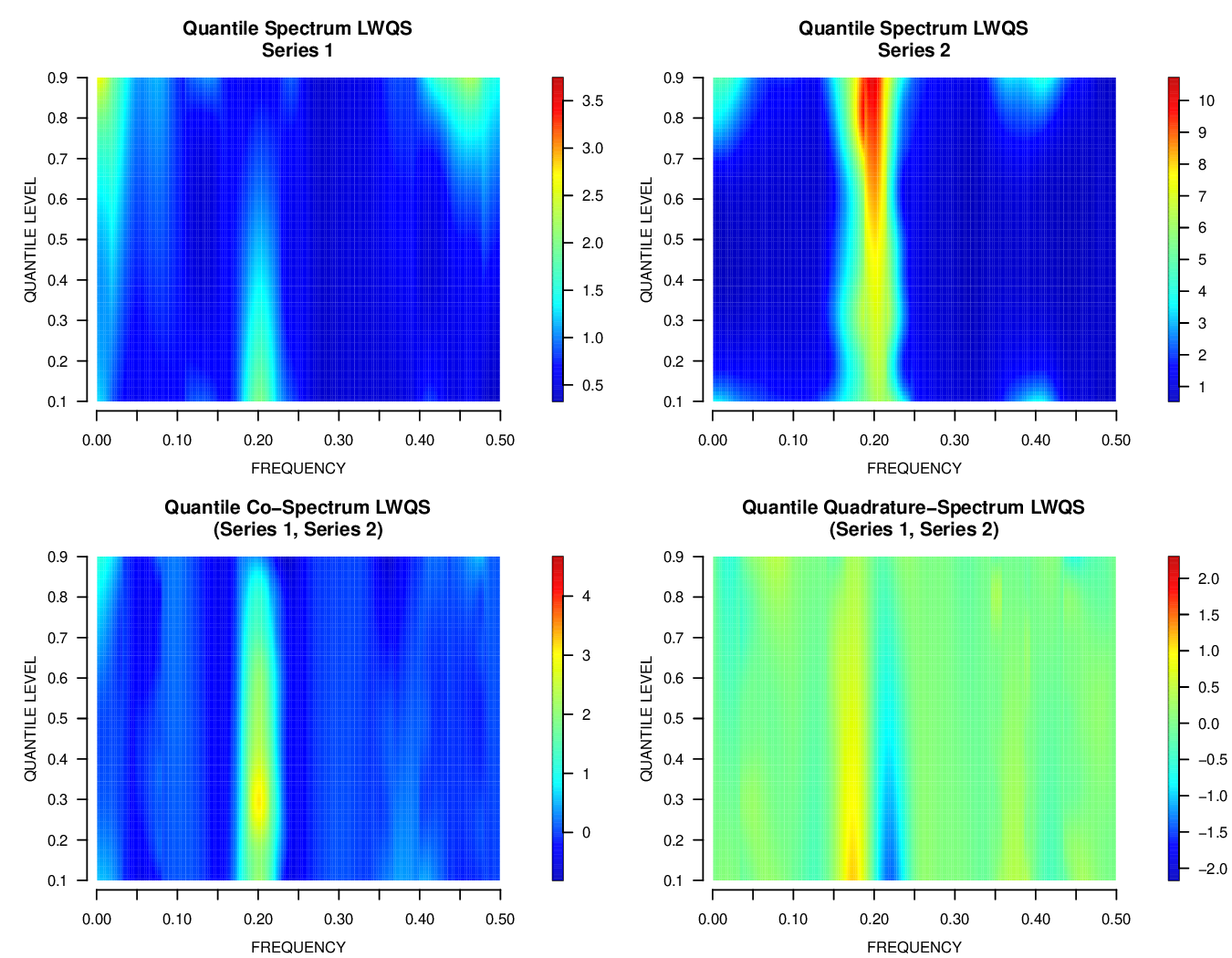}  
\caption{LWQS estimates of the quantile spectrum and cross-spectrum shown in Figure~\ref{fig:qspec} from the series shown in Figure~\ref{fig:ts} using {\tt gamm} with correlated residuals.  KLD = 0.130.}
 \label{fig:lwqsgamm}
\end{figure}

Figure~\ref{fig:ts} shows the series from one of the simulation runs. The corresponding quantile periodogram and cross-periodogram are shown in Figure~\ref{fig:qcper}. Figure~ \ref{fig:qser} 
depicts the QSER  at $\al=0.1$, $0.5$, and $0.9$. The corresponding QACFs are shown in Figure~\ref{fig:qacf}.

Figure~\ref{fig:lw} shows the LW estimates obtained from the series in Figure~\ref{fig:ts}.
These  estimates are constructed according to (\ref{LW}) using the Tukey-Hanning window (\ref{Hanning}) with $M=30$. They can be viewed as a smoothed version of the quantile periodogram and cross-periodogram  in  Figure~\ref{fig:qcper} with respect to the frequency variable.

Figure~\ref{fig:lwqs} shows the LWQS estimates obtained by applying quantile smoothing 
to the LW estimates in  Figure~\ref{fig:lw} using  the R function {\tt smooth.spline} with
the smoothing parameter chosen by the generalized cross-validation (GCV) criterion (R Core Team 2024). 
The resulting KLD equals 0.194, which is reduced slightly from 0.198 achieved by the LW estimates  in  Figure~\ref{fig:lw}. The visual effect of quantile smoothing is not quite noticeable when compared to  Figure~\ref{fig:lw}.  

A better result is shown in Figure~\ref{fig:lwqs2}. These estimates are also obtained by 
{\tt smooth.spline}, but the smoothing parameter {\tt spar} is set to 0.9 instead of being determined by GCV.
The estimates in Figure~\ref{fig:lwqs2} appear less noisy  when compared to the estimates in Figures~\ref{fig:lw} and \ref{fig:lwqs}. The KLD is reduced significantly to 0.109 from 0.198 and 0.194, respectively.

A closer examination of the LW estimates reveals strong positive correlations across quantiles.
Such correlations are also suggested by the theoretical analysis of quantile regression solutions 
generally in the IID case (Koenker 2005, p.\ 72).  The {\tt smooth.spline} function with GCV is 
not designed to handle correlated data effectively due to the underlying assumption
of IID errors. To take the correlations into account, we  use the R function {\tt gamm} 
in the `mgcv' package (Wood 2022). Under the framework 
of generalized additive mixture models (Wang 1998), this function accommodates the correlations 
as a random effect. It jointly estimates the parameters of a user-specified correlation structure 
and the parameters of the smoothing splines while using GCV for smoothing parameter selection.

\begin{figure}[p]
\centerline{\includegraphics[height=3in,angle=-90]{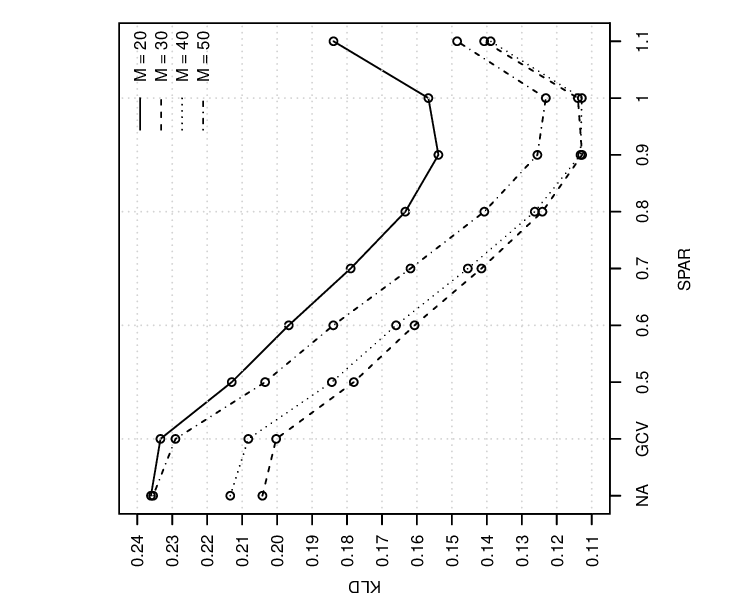} } 
\vspace{-0.3in}
\caption{Mean KLD of the LWQS estimator for the mixture process (\ref{y1})-(\ref{y2}) with different bandwidth parameter $M$ 
when quantile smoothing is performed by {\tt smooth.spline} with different smoothing parameter {\tt spar} (NA for no quantile smoothing). Results are based on 1000 Monte Carlo runs.} 
\label{fig:mse}
\bigskip\bigskip
{
\begin{center} 
\caption*{Table 1: Mean KLD of the LWQS Estimator for the Mixture Process (\ref{y1})-(\ref{y2})}
\begin{tabular}{lrrrrrrrr} \hline
\multicolumn{1}{l}{Quantile Smoothing Method} & 
$M=20$ &  $M=30$ & $M=40$ &  $M=50$   \\ \hline
no quantile smoothing & 0.236  & 0.204 &  0.213  & 0.235    \\
 {\tt smooth.spline} with GCV & 0.233  & 0.200  &  0.208  & 0.229   \\
 {\tt smooth.spline} with {\tt spar} = 0.9 & 0.154   & 0.113  &  0.113 & 0.126    \\
 {\tt gamm} with correlated residuals &  0.178  & 0.137  & 0.139  & 0.153  \\ \hline
\end{tabular} 
\end{center}
} 
\centerline{\scriptsize \hspace{-2.9in}Results are based on 1000 Monte Carlo runs.}
\end{figure}

Figure~\ref{fig:lwqsgamm} shows the result of {\tt gamm} that employs 
the AR(1) correlation structure as an surrogate (true correlations
are not available because they depend on the unknown spectrum). The KLD of these estimates equals 0.130, a significant improvement over  {\tt smooth.spline} with GCV. This improvement is achieved at a higher computational cost: a 100-fold increase in computing time when compared to {\tt smooth.spline}.  

Based on 1000 Monte Carlo runs, Figure~\ref{fig:mse} and Table~1 provide  a more comprehensive assessment 
of the LWQS estimator using {\tt smooth.spline} and {\tt gamm} for quantile smoothing.
As shown in Figure~\ref{fig:mse}, {\tt smooth.spline} with GCV offers a slight improvement over 
no quantile smoothing;  a significant improvement can be made by setting {\tt spar} 
manually within a range of values, with the best choice  being somewhere between 0.9 and 1.0. 
The results shown in  Table~1 confirm the superiority of {\tt gamm} with correlated residuals over {\tt smooth.spline} when the smoothing parameter is selected automatically by GCV.

\bigskip
\noindent
{\it Estimation of Quantile Spectrum at Single Quantile Levels}

\smallskip
For some applications, it is sufficient to treat the quantile spectrum  as a 
function of frequency at a single quantile level  rather than a bivariate function of frequency
and quantile level on $[0,\pi) \times (0,1)$. This restricted view is typically held in the conventional 
spectral analysis of indicator processes (e.g., Davis and Mikosch 2009; Hagemamm 2013; 
Dette et al.\ 2015; Barun\'{i}k and Kley 2019)).  For the quantile spectrum defined by (\ref{S}), 
an interesting question  is whether or not the quantile smoothing techniques for estimating the quantile spectrum as a bivariate function remains beneficial to the estimation of quantile spectrum at the single quantile level of interest. Intuitively, the answer to this question should be affirmative when the reduction of statistical variability 
achieved by quantile smoothing outweighs the increase of bias it introduces to the estimate. A typical requirement for this benefit to materialize is that the quantile spectrum should vary smoothly with $\al$ in 
a neighborhood of the quantile level of interest.

Our next experiment is intended to evaluate this  benefit for the mixture process (\ref{y1})-(\ref{y2}).
We consider estimating the quantile spectrum in Figure~\ref{fig:qspec} as functions of frequency 
at three quantile levels: $\al = 0.3, 0.5, 0.8$. We obtan the LWQS estimates of these spectra by applying
a quantile smoothing procedure to the LW estimates on the entire grid of quantile levels 
as in Table 1, and then extracting the smoothed estimates at each quantile level of interest. 
The mean KLD of the resulting estimates are calculated from 1000 Monte Carlo runs and shown in Table 2.
This result clearly supports the assertion that  under the smoothness condition treating the quantile spectrum 
as a bivariate function with suitable quantile smoothing can improve the accuracy for estimating 
the quantile spectrum at a single quantile level.

\begin{table}[t]
{
\begin{center} 
\caption*{Table 2: Mean KLD of the LWQS Estimator for Estimating the Quantile Spectrum 
of the Mixture Process (\ref{y1})-(\ref{y2}) at Single Quantile Levels}
\begin{tabular}{lcrrrrrr} \hline
\multicolumn{1}{l}{Quantile Smoothing Method} & $\al$ &
$M=20$ &  $M=30$ & $M=40$ &  $M=50$   \\ \hline
no quantile smoothing                           & 0.3 & 0.284  & 0.219 &  0.219  & 0.237  \\
 {\tt smooth.spline} with GCV                 & 0.3 & 0.282  & 0.216 &  0.214  & 0.231  \\
 {\tt smooth.spline} with {\tt spar} = 0.9 & 0.3 & 0.190  & 0.114 &  0.102  & 0.108  \\
 {\tt gamm} with correlated residuals    &  0.3 & 0.217 & 0.141  &  0.131  & 0.141  \\ \hline
 no quantile smoothing                          & 0.5 & 0.262  & 0.208 &  0.211  & 0.231  \\
 {\tt smooth.spline} with GCV                 & 0.5 & 0.260  & 0.205 &  0.207  & 0.225  \\
 {\tt smooth.spline} with {\tt spar} = 0.9 & 0.5 & 0.169  & 0.110 &  0.106  & 0.118  \\
 {\tt gamm} with correlated residuals    &  0.5 & 0.209 & 0.144  & 0.136   & 0.145  \\ \hline
 no quantile smoothing                          & 0.8 & 0.162  & 0.178 &  0.203  & 0.231  \\
 {\tt smooth.spline} with GCV                 & 0.8 & 0.159  & 0.174 &  0.198  & 0.224  \\
 {\tt smooth.spline} with {\tt spar} = 0.9 & 0.8 & 0.081  & 0.085 &  0.100  & 0.118  \\
 {\tt gamm} with correlated residuals    &  0.8 & 0.099 & 0.105  &  0.121  & 0.138  \\
\hline
\end{tabular} 
\end{center}
}
\centerline{\scriptsize \hspace{-3.3in}Results are based on 1000 Monte Carlo runs.}
\end{table}

\bigskip
\noindent
{\it Additional Simulation Results}

\smallskip
To validate the findings from the first set of simulated data, we repeat the experiments with 
the second set of simulated data which is generated from an ARMA process $\by_t := [y_{1,t},y_{2,t}]^T$:
\eqn
\by_t - \bA_1 \, \by_{t-1} - \bA_2 \, \by_{t-2} =  \bmep_t + \bB \, \bmep_{t-1},  \quad  
\{\bmep_t \} \sim \text{ IID } \N(\0,\bSig),
\label{arma21}
\eqqn
where 
\eq
\bA_1 & := & \left[
\begin{array}{rr}
0.816 & 1.246 \\
0.558 & 1.107
\end{array} 
\right], \quad 
\bA_2 := \left[
\begin{array}{rr}
0.643 & 1.184 \\
0.307 & 0.203
\end{array} 
\right], \\
\bB & := & \left[
\begin{array}{rr}
0 & 2.496 \\
0.4 & 0
\end{array} 
\right], \quad
 \bSig := \left[
\begin{array}{rr}
0.04 & -0.02 \\
-0.02 & 0.02
\end{array} 
\right].
\eqq
The results for this  data set are provided in Appendix I. All findings from the first set of simulated data 
are confirmed by these results.

\newpage
\bigskip
\noindent
{\it Yearly Sunspot Numbers}

\begin{figure}[p]
\centering
\includegraphics[height=3in,angle=-90]{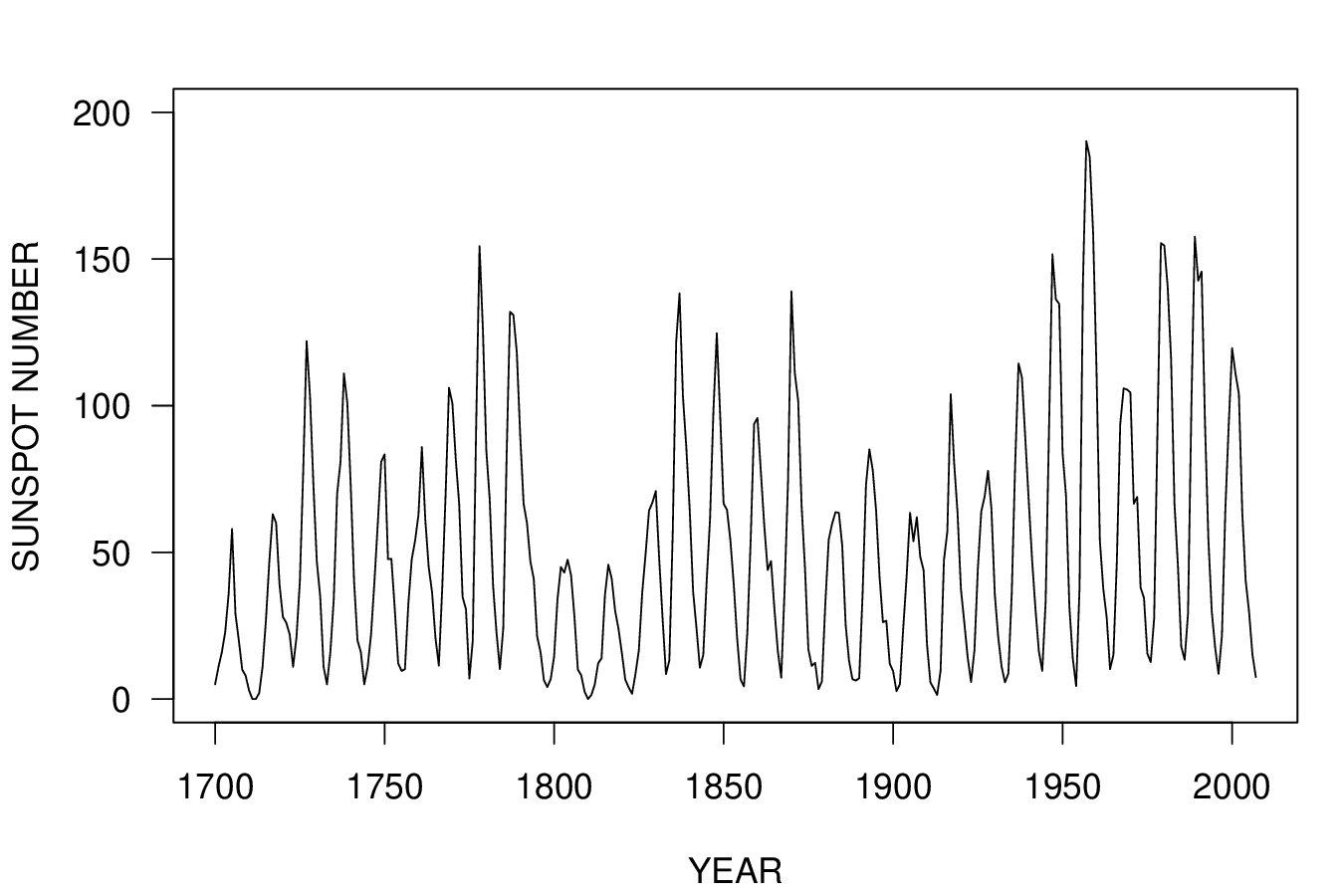} 
\includegraphics[height=3in,angle=-90]{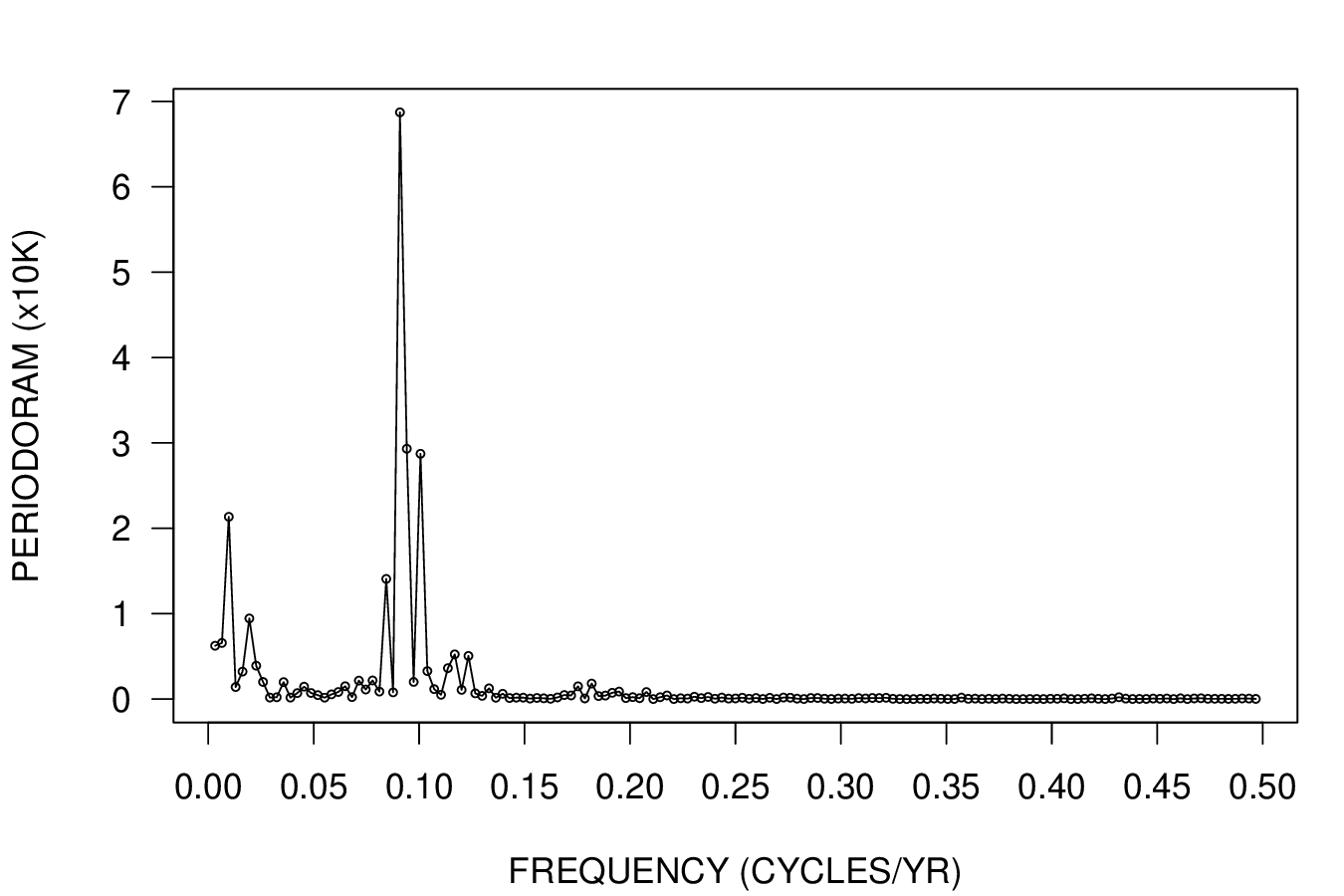} 
\caption{Time series of yearly sunspot numbers $(n=308)$ and  its periodogram.}
\label{fig:sunspot}
\vspace{0.4in}
\includegraphics[height=3in,angle=-90]{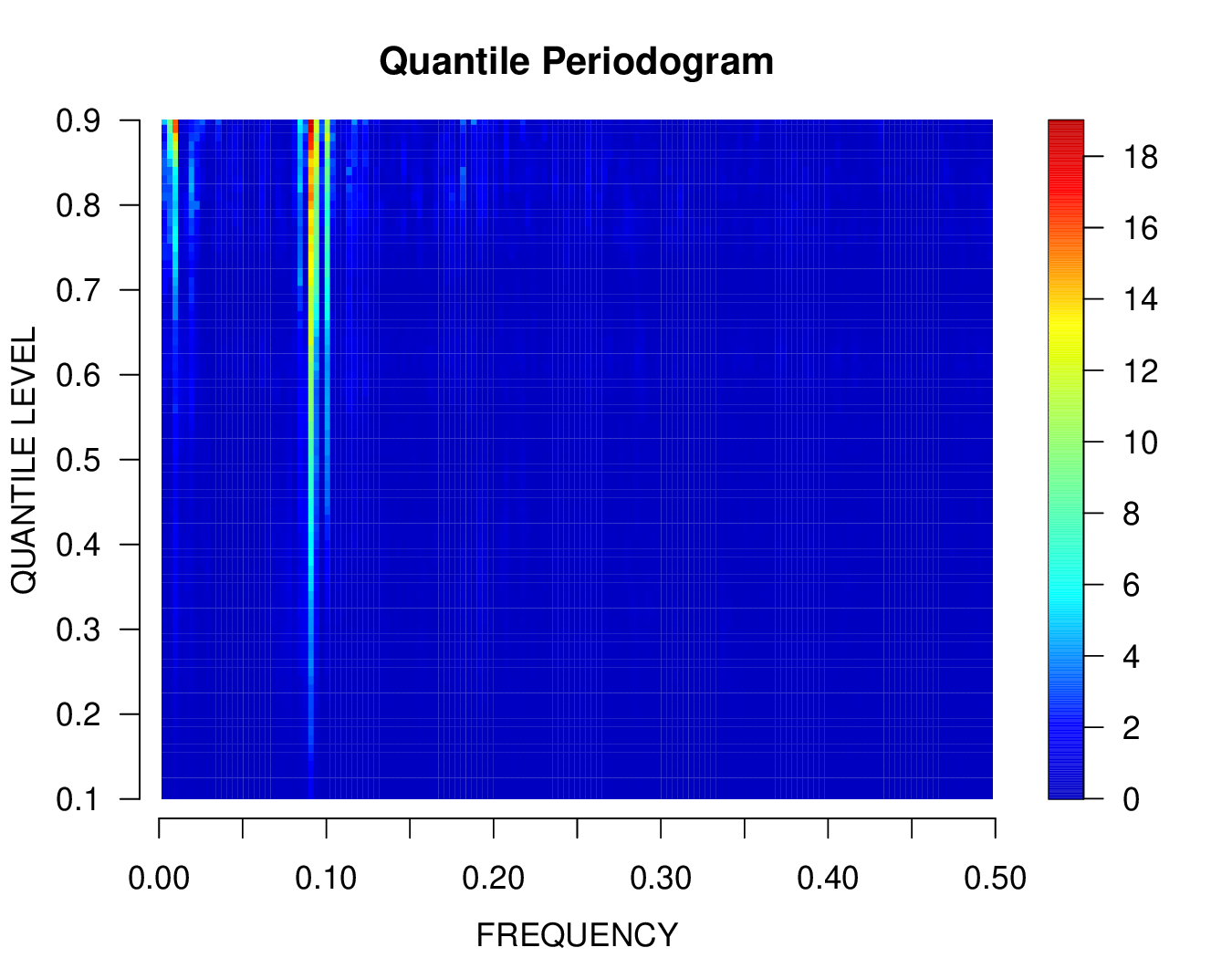}  
\includegraphics[height=3in,angle=-90]{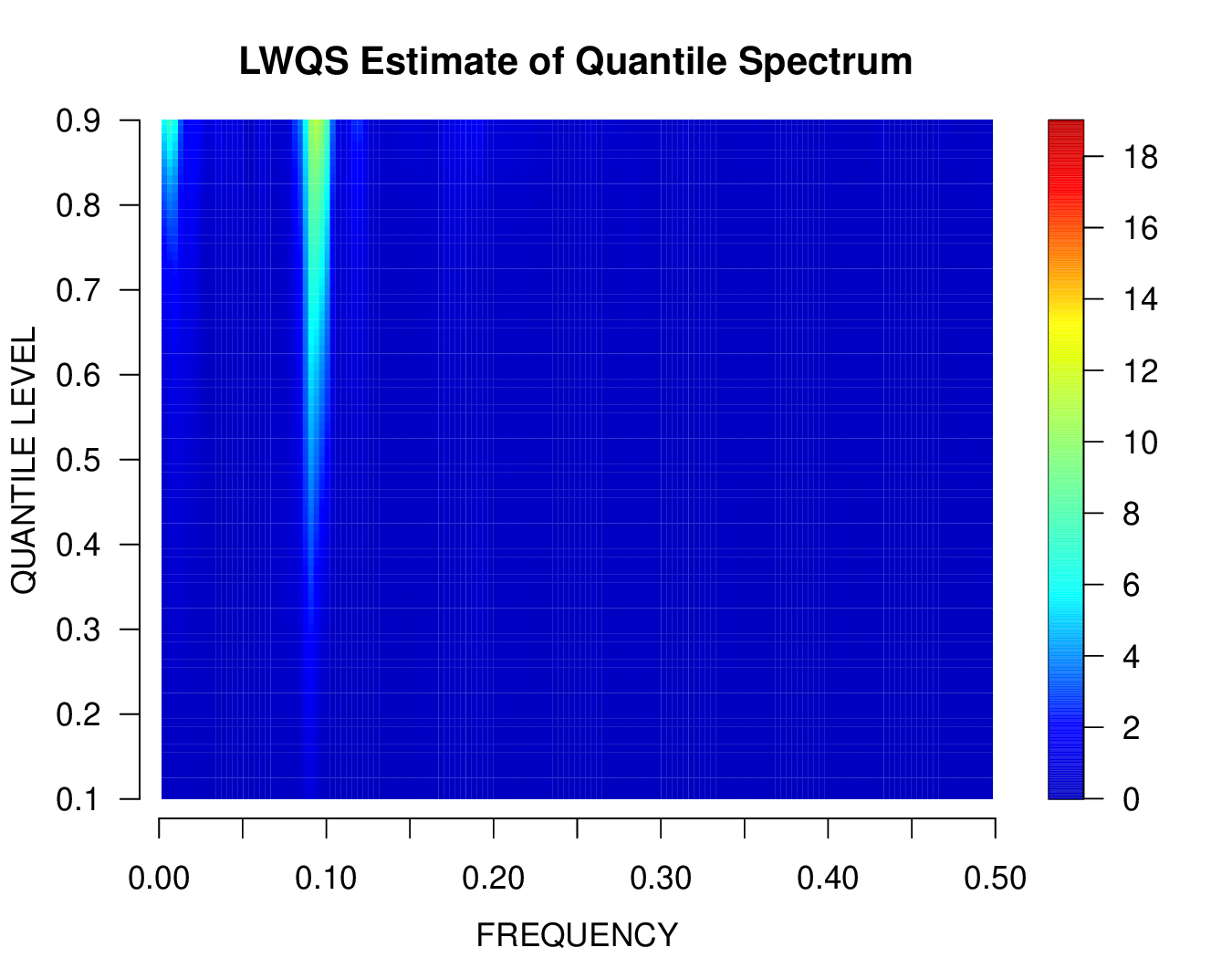}  
\caption{Quantile periodogram (left) and LWQS spectral estimate (right) of the sunspot numbers obtained
with $M=150$ and {\tt smooth.spline} ({\tt spar}=0.9). }
 \label{fig:qper:sunspot}
\end{figure}

Finally, we present an real-data example. It is a time series  consisting of $n=308$
yearly sunspot numbers from year 1700  to 2007 (Li 2012). As shown in Figure~\ref{fig:sunspot}, 
this series exhibits a strong 11 year cycle ($\om_{28} = 2\pi \times 28/308$)  
and a weaker 102.7 year cycle  ($\om_{3} = 2\pi \times 3/308$)  which can be seen in the ordinary periodogram.

The quantile periodogram of this series on the quantile grid $\{0.10,0.11,\dots,0.90\}$ 
 is shown in Figure~\ref{fig:qper:sunspot} together with the LWQS spectral estimate 
 obtained using the Tukey-Hanning window (\ref{Hanning}) with $M=150$; quantile smoothing
 is performed by {\tt smooth.spline} with {\tt spar} = 0.9. As noted in Li (2012), 
 the quantile periodogram is able to depict an additional property of the two spectral peaks: i.e.,
 their magnitude is reduced continuously  as the quantile level decreases from 0.9 to 0.1.
 The spectral peaks and their decreasing trend, unavailable in the ordinary periodogram, are shown more clearly in 
 the LWQS estimate, thanks to the smoothing effect of this estimate in both frequency and quantile level.

\begin{figure}[t]
\centering
\includegraphics[height=3in,angle=-90]{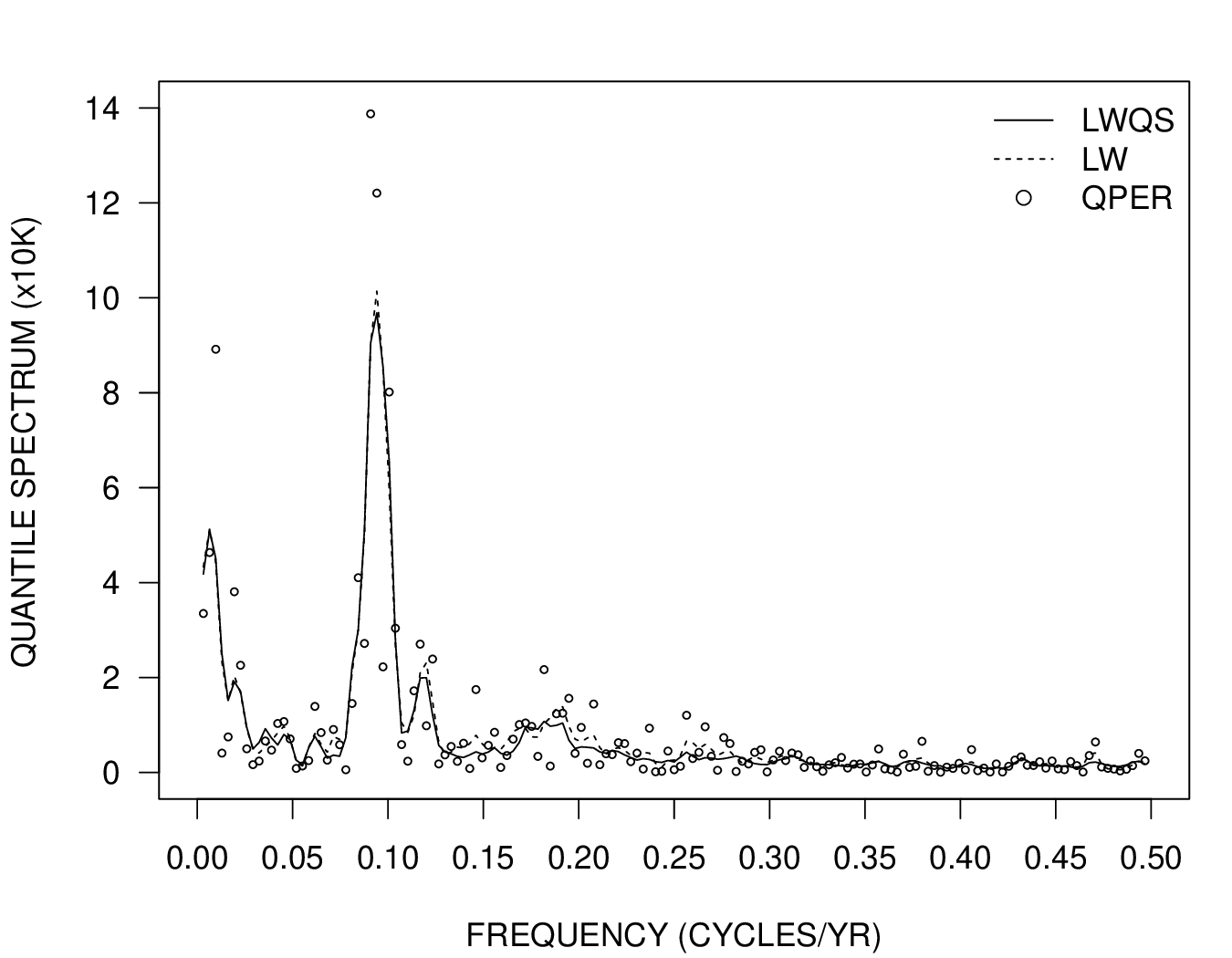} 
\includegraphics[height=3in,angle=-90]{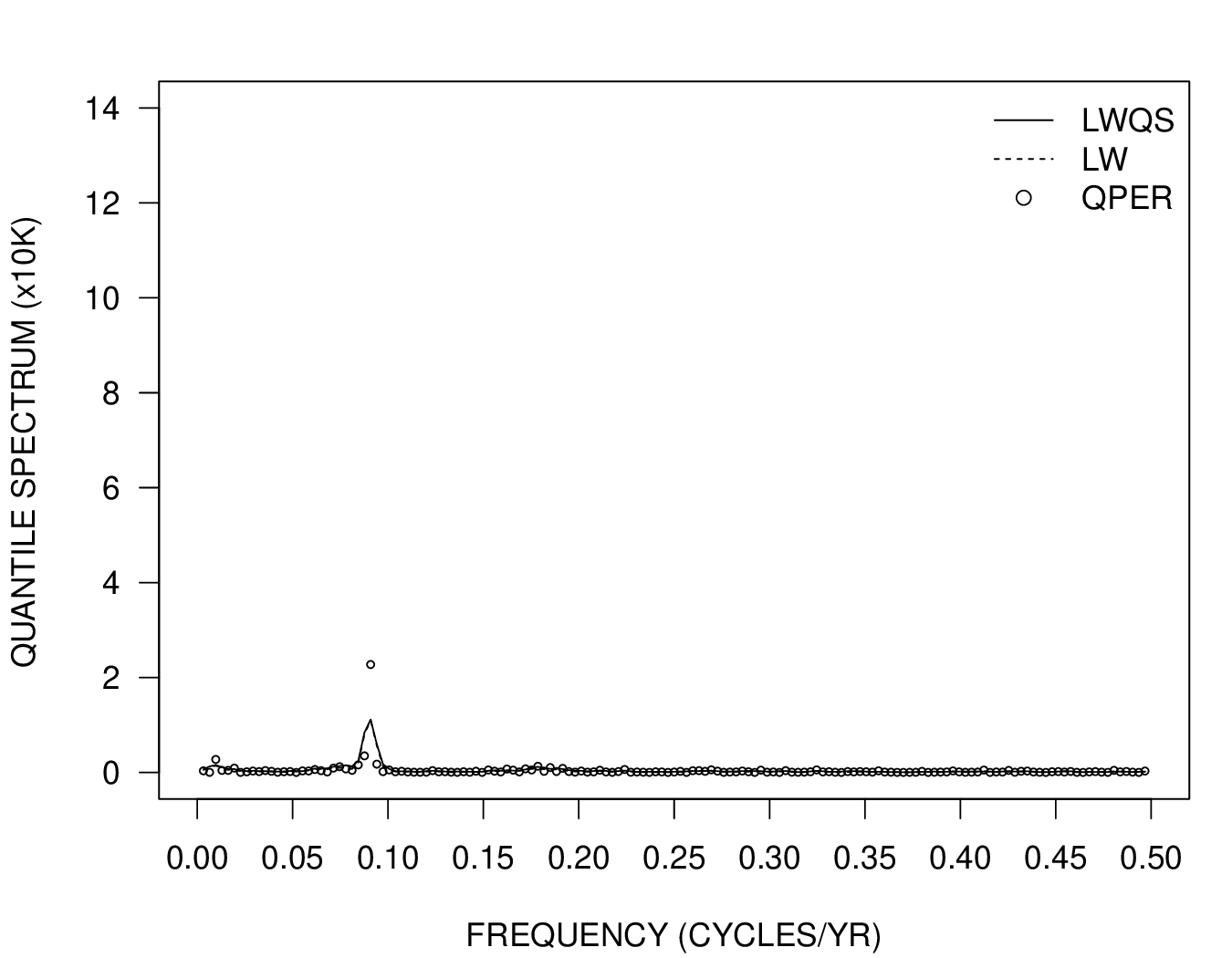} 
\caption{Quantile periodogram and LWQS spectral estimate of the sunspot numbers at quantile 
level $\al=0.85$ (left) and $\al=0.15$ (right).}
\label{fig:qper:sunspot2}
\end{figure}

The smoothing effect can be further appreciated by inspecting Figure~\ref{fig:qper:sunspot2},
where  the quantile periodogram and the LWQS estimate are plotted at selected quantile levels as 
functions of frequency.  The LWQS estimate is able to reduce the noise in the quantile periodogram 
significantly so that the spectral peaks are better delineated. Quantile smoothing in LWQS 
offers further reduction of background noise in the LW estimate.

\section{Concluding Remarks}

In this paper, we propose a nonparametric method for estimating the quantile spectrum
introduced  in Li (2008; 2012; 2014)  through trigonometric quantile regression. 
This method is based on the quantile discrete Fourier transform (QDFT) by the trigonometric 
quantile regression and the quantile series (QSER) as the inverse Fourier transform of the QDFT. 
The autocovariance function  of the QSER, or QACF, enables the construction 
of a lag-window (LW) estimator for the quantile spectrum. We also employ a smoothing 
procedure to smooth the LW estimates across quantiles in order to further improve the estimation 
accuracy when the underlying spectrum is smooth with respect to the quantile level.  
Quantile smoothing is shown in our simulation study 
to be effective not only for estimating the quantile spectrum as a bivariate function of frequency
and quantile level but also as a univariate function of frequency for a fixed quantile level.

The spline smoothing method implemented by the R function {\tt gamm}
turns out to be quite effective in our simulation study. An AR(1)-type correlation function is employed 
as a surrogate for the correlations of the LW estimates
across quantiles. Whether it can be improved by a more suitable correlation function remains
a topic for future research. Furthermore, smoothing across quantiles for each fixed frequency, 
as done in this paper, may introduce artifacts of discontinuity across frequencies. To overcome this difficulty,
it is helpful to develop  an alternative method that has the effect of smoothing jointly
across frequencies and quantile levels.

The nonparametric method in this paper complements the semi-parametric method
investigated in Chen et al.\ (2019) and Jim\'{e}nez-Var\'{o}n et al.\ (2024). The latter derives 
a parametric autoregressive (AR) model from the quantile periodogam at fixed quantile levels 
and  then smooths the AR parameters across quantiles by a nonparametric smoother. 
With the help of the QSER introduced in this paper, it becomes possible to fit the AR model 
directly to the QSER by least squares or through the QACF by solving the Yule-Walker equations.
A comprehensive treatment of this approach will be the topic of  a future paper. 

Finally, the idea of deriving QSER through QDFT by trigonometric quantile regression 
can be generalized to other objective functions. An example is the $M$-periodogram 
discussed in Fajardo et al.\ (2018), where Huber's $\psi$-function and the $\ell_p$ norm for $p \in (1,2)$ are employed in place of quantile regression. In addition to periodogram smoothing, an LW estimator can be obtained from a suitably
defined time-domain representation analogous to QSER.

\section*{References}

{\footnotesize
\begin{description} 

\item
Brockwell, P., and Davis, R. (1991) {\it Time Series: Theory and Methods},
2nd edn, section 11.6. New York: Springer.

\item
Barun\'{i}k, J., and Kley, T. (2019) Quantile coherency: A general measure for dependence between
cyclical economic variables. {\it Econometrics Journal}, 22, 131--152.

\item
Chen, T., Sun, Y., and Li, T.-H. (2019) A semiparametric estimation algorithm for the 
quantile spectrum with an application to earthquake classification using convolutional neural network. 
 {\it Computational Statistics \& Data Analysis}, 154, 107069.

\item
Davis, R., and Mikosch, T. (2009) The extremogram: A correlogram for extreme events. 
{\it Bernoulli}, 15, 977--1009.

\item 
Dette, H., Hallin, M., Kley, T., and Volgushev, S. (2015) Of copulas,
quantiles, ranks and spectra: an $L_1$-approach to spectral analysis. 
{\it Bernoulli}, 21, 781--831.

\item
Fajardo, F., Reisen, V., L\'{e}vy-Leduc, C., and Taqqu, M.\ (2018)
$M$-periodogram for the analysis of long-range-dependent time series.
{\it Statistics}, 52, 665--683.

\item
Hagemann, A. (2013) Robust spectral analysis. arXiv:1111.1965.

\item
Jim\'{e}nez-Var\'{o}n, C., Sun, Y., and Li, T.-H. (2024)
A semi-parametric estimation method for  quantile coherence  with an application to bivariate financial time series clustering. {\it Econometrics and Statistics}. \url{https://doi.org/10.1016/j.ecosta.2024.11.002}.

\item
Kakizawa, Y., Shumway, R., and Tanaguchi, M. (1998) Discrimination and clustering for multivariate time series. {\it  Journal of the American Statistical Association}, 93, 328--340.

\item
Koenker, R. (2005) {\it Quantile Regression}. Cambridge, UK: Cambridge University Press.

\item
Li, T.-H. (2008)  Laplace periodogram for time series analysis.
 {\it Journal of the American Statistical Association}, 103, 757--768. 

\item
Li, T.-H. (2012) Quantile periodograms. {\it Journal of the American Statistical Association}, 107, 765--776. 

\item 
Li, T.-H. (2014) {\it Time Series with Mixed Spectra}. Boca Raton, FL: CRC Press.

\item
Li, T.-H. (2020) From zero crossings to quantile-frequency analysis of time
series with an application to nondestructive evaluation. 
{\it Applied Stochastic Models for Business and Industry}, 36, 1111--1130

\item
Li, T.-H. (2021) Quantile-frequency analysis and spectral measures for diagnostic checks of time series with nonlinear dynamics. {\it Journal of the Royal Statistical Society Series C: Applied Statistics}, 70, 270-290.

\item
Li, T.-H. (2023) Quantile-frequency analysis and deep learning for signal classification.
{\it  Journal of Nondestructive Evaluation},  42, DOI:10.1007/s10921-023-00952-y.

\item
Percival, D., and Walden, A. (1993) {\it Spectral Analysis for Physical Applications}. 
Cambridge, UK: Cambridge University Press.

\item
Portnoy, S. (1991) Asymptotic behavior of the number of regression quantile breakpoints. 
{\it SIAM journal on scientific and statistical computing}, 12, :867--883.

\item
Priestley, M. (1981) {\it Spectral Analysis and Time Series}, p.~443. New York: Academic Press.

\item
R Core Team (2024) R: A language and environment for statistical
  computing. R Foundation for Statistical Computing, Vienna,
  Austria. \url{https://www.R-project.org/}.

\item
Wang, Y. (1998) Smoothing spline models with correlated random errors.
{\it Journal of the American Statistical Association}, 93, 341--348.

\item
Whittle, P. (1953) Estimation and information in stationary time series. {\it Arkiv f\"{o}r Matematik}, 2, 423--434.

\item
Wood, S. (2022) Package `mgcv'. \url{https://cran.r-project.org/web/packages/mgcv/mgcv.pdf}.

\item
Wu, W. (2007) $M$-estimation of linear models with dependent errors. {\it Annals of Statistics}, 35, 495--521.

\end{description}
}

\newpage

\section*{Appendix II: The R Functions}

The following is a summary of the relevant R functions for the proposed method. 
These functions are available in the R package `qfa' (version $\ge 3.1$) which can be installed from 
\url{https://cran.r-project.org}. An installable R package  is also available at \url{https://github.com/thl2019/QFA}.

\begin{itemize}
\item {\tt qdft}: a function that computes the quantile discrete Fourier transform (QDFT) of a univariate or multivariate time series at a user-specified sequence of quantile levels.
\item {\tt qser}: a function that computes the quantile series (QSER) of a 
univariate or multivariate time series at a user-specified sequence of quantile levels from the time series 
or the QDFT produced by {\tt qdft}.
\item {\tt qacf}: a function that computes the quantile autocovariance function (QACF) of a 
univariate or multivariate time series at a user-specified sequence of quantile levels from the time series 
or the QDFT produced by {\tt qdft}.
\item {\tt qspec.lw}: a function that computes the lag-window spectral estimate with or without quantile smoothing (LWQS or LW) for a given bandwidth parameter from the QACF produced by {\tt qacf}.
\item {\tt qdft2qser}: a function that computes the quantile series (QSER) from the QDFT produced by {\tt qdft}.
\item {\tt qdft2qper}: a function that computes the quantile periodogram 
 (QPER) from the QDFT produced by {\tt qdft}.
\item {\tt qdft2qacf}: a function that computes the quantile autocovariance function (QACF) 
from the QDFT produced by  {\tt qdft}.
\item {\tt tqr.fit}: a low-level function that computes the trigonometric quantile regression (TQR) solution 
for a single frequency at a user-specified sequence of quantile levels.
\end{itemize}

\end{document}

%% file: myfonts.tex
\DeclareMathAlphabet{\mybm}{OT1}{ptm}{b}{it}

\usepackage{mathptmx}
\usepackage{euscript}
\usepackage{latexsym}
\usepackage{amsmath}
\usepackage{amsfonts}
\usepackage{amssymb}
\usepackage{ulem}
\usepackage{amsthm}

{\theoremstyle{remark} 
}

{\theoremstyle{definition} 

}

\newcommand{\BIBO}{{\small \BIBO}}

\newcommand{\dlim}{\stackrel{\raisebox{-0.06in}{$\scriptscriptstyle D$\,}}{\rightarrow}}

\newcommand{\linfty}{\rightarrow \infty}
\newcommand{\lzero}{\rightarrow 0}

\newcommand{\eq}{\begin{eqnarray*}}
\newcommand{\eqq}{\end{eqnarray*}}
\newcommand{\eqn}{\begin{eqnarray}}
\newcommand{\eqqn}{\end{eqnarray}}

\newcommand{\eqb}{\begin{align*}}
\newcommand{\eqqb}{\end{align*}}

\newcommand{\tr}{\mathrm{tr}}
\newcommand{\N}{\mathrm{N}}

\newcommand{\eqna}{\begin{align}}
\newcommand{\eqqna}{\end{align}}

\newcommand{\bA}{\mathbf{A}}
\newcommand{\bB}{\mathbf{B}}

\newcommand{\bQ}{\mathbf{Q}}

\newcommand{\bS}{\mathbf{S}}

\newcommand{\bx}{\mathbf{x}}
\newcommand{\by}{\mathbf{y}}

\newcommand{\bbR}{\mathbb{R}}

\newcommand{\0}{{\bf 0}}

\newcommand{\al}{\alpha}

\newcommand{\ep}{\epsilon}
\newcommand{\gam}{\gamma}

\newcommand{\lam}{\lambda}

\newcommand{\om}{\omega}

\newcommand{\bmep}{\pmb{\epsilon}}

\newcommand{\bmzeta}{\pmb{\zeta}}

\newcommand{\bGam}{\pmb{\Gamma}}

\newcommand{\bSig}{\pmb{\Sigma}}

